 \documentclass[11pt,twocolumn]{article}

 \pdfoutput=1

 \usepackage[final]{pdfpages} 

\usepackage[round]{natbib}   

\usepackage[english]{babel}

\usepackage{graphicx}

\usepackage{amsmath}
\usepackage{chemformula} 

\usepackage{enumitem}

\usepackage{color}

\usepackage{float}
\usepackage{ifthen}

\usepackage{hyperref}

\usepackage{titlesec}

\titlespacing*{\section}{0pt}{1.1\baselineskip}{\baselineskip}

\begin{document}

\def\hmath#1{\text{\scalebox{1.5}{$#1$}}}
\def\lmath#1{\text{\scalebox{1.4}{$#1$}}}
\def\mmath#1{\text{\scalebox{1.2}{$#1$}}}
\def\smath#1{\text{\scalebox{.8}{$#1$}}}

\def\hfrac#1#2{\hmath{\frac{#1}{#2}}}
\def\lfrac#1#2{\lmath{\frac{#1}{#2}}}
\def\mfrac#1#2{\mmath{\frac{#1}{#2}}}
\def\sfrac#1#2{\smath{\frac{#1}{#2}}}

\def\pow{^\mmath}



\twocolumn[

\begin{center}
{\bf \Large {The absolute seawater entropy: Part~II. Case studies}}
\\
\vspace*{3mm}
{\Large by Dr.Hab. Pascal Marquet}. \\
\vspace*{3mm}
{\large Retired from M\'et\'eo-France (CNRM), Toulouse, France.}
\\ \vspace*{2mm}
{\large  \it E-mail: pascalmarquet@yahoo.com}
\\ \vspace*{2mm}
{\large URL: \url{https://sites.google.com/view/pascal-marquet}}
\\ \vspace*{2mm}
{\Large\bf  \color{black} Accepted in:}
{\Large\bf  \color{black} \it\bf 
{Comptes Rendus Geoscience (Paris, France),}}
\vspace*{2mm}\\
{\large  \color{black} {Initial submission: 
                       7th of October 2024}.}
\vspace*{1mm}\\
{\large  \color{black} Three Reviewers' comments received 
                       the 10th of September, 2025.}
\vspace*{1mm}\\
{\large  \color{black} First revised version: 
                       30th of November, 2025.} 
\vspace*{1mm}\\
{\large  \color{black} Second Reviewers' comments received 
                       the 11th of February, 2026.}
\vspace*{1mm}\\
{\large  \color{black} {\bf Accepted version:
                       25th of March, 2026} \;\;\;
                 \url{https://doi.org/10.5802/crgeos.334}.} 
\end{center}

\begin{center}
{\large \bf Abstract}
\end{center}
\vspace*{-3mm}

\hspace*{7mm}
The aim of this second part of the article is to study the absolute definition of the seawater entropy described in Part I with several concrete cases. 
Observed vertical profiles and polar transects, as well as analysed surface data, show that very different temperatures and salinity values can organise to create new isentropic regions. 
This can only be revealed by the absolute formulation of the entropy of seawater (Arctic Ocean; Bay of Bengal; Mediterranean, Black, and Caspian Seas).
Existing hypotheses to explain these results include the possible impact of turbulent processes that must be applied to the entropies of the atmosphere and oceans.
\vspace*{0mm}

\begin{center}

First-Review Answers to the Editors and Reviewers 
can be found on Zenodo \citep{Marquet_Zenodo_2025_Answers_R1}.
\vspace*{2mm}

Supplementary materials are provided in the Zenodo file
\citet{Marquet_Zenodo_2025_Sup_Mat_3rd_law}. 
\vspace*{2mm}

Second-Review Answers to the Reviewers 
can be found on Zenodo \citep{Marquet_Zenodo_2026_Answers_R2}.
\vspace*{2mm}
\end{center}

%
] 

 \section{Introduction}
\label{section_introduction}
\vspace*{-2mm}

The present Part-II paper follows on from the Part-I paper \citep{Marquet_CRAS_Geos_25_I}, 
in which the absolute entropy of the ocean $\eta_{\rm abs}$ (expressed in J~K${}^{\,-1}$~kg${}^{\,-1}$) 
was defined as a modification of the standard arbitrary TEOS10 version $\eta_{\rm std/TEOS10}$  
\citep[see][]{Feistel_TEOS_manual_2010}, according to
\begin{align}
\eta_{\rm abs} & = \eta_{\rm std/TEOS10}  \;+\; \Delta \eta_{\rm s}  
\label{Eq_Delta_eta_ans_std} 
\; , \\
\Delta \eta_{\rm s} & = (\eta_{\rm s0}-\eta_{\rm w0}) \times 
\frac{(S_{\rm A}-S_{\rm SO})}{1000} 
\label{Eq_etas_minus_etaw} 
\; , \\
 \eta_{\rm w0} & \: \approx 3513.4 
 \pm 1.7~\mbox{J~K${}^{\,-1}$~kg${}^{\,-1}$}
 \label{eq_eta_w0_NEA_TDB_1992} 
  \;\;\mbox{(at $0{}^{\circ}$C)} 
\; , \\
\eta_{\rm s0} & \approx
    1633.3 \pm 15~\mbox{J~K${}^{\,-1}$~kg${}^{\,-1}$}
  \;\;\mbox{(at $0{}^{\circ}$C)} 
 \label{eq_eta_s0_NEA_TDB_1992} 
\; , \\
\Delta \eta_{\rm s} 
& \approx (-1880 \pm 17)
\times \frac{(S_{\rm A}-S_{\rm SO})}{1000} 
\label{Eq_etas_minus_etaw_value} 
\; ,
\end{align}
where $\eta_{\rm w0}$ and $\eta_{\rm s0}$ are the absolute entropies
for pure water and sea salts (respectively), 
$S_{\rm A}$ is the absolute salinity (in g~kg${}^{\,-1}$), 
and $S_{\rm SO}=35.165\,04$~g~kg${}^{-1}$ is the standard value for which the 
standard TEOS10 seawater entropy is cancelled at the ambiant temperature 
$t_{\rm S0}=0$°C by using arbitrary tuning values for both 
$\eta_{\rm w0}$ and $\eta_{\rm s0}$.

The aim of this second part is to demonstrate the physical meaning of this new 
absolute definition $\eta_{\rm abs}$ by studying the impacts of the salinity increment 
$\Delta \eta_{\rm s}$ on observed oceanic vertical profiles and analysed datasets.
It is shown that the absolute version of the seawater entropy allows the observed 
data to be described in a new light. 
In particular, new isentropic regions both vertically and horizontally can only 
be revealed by the absolute and thermodynamically consistent definition of 
the seawater entropy.

The paper is organised as follows.
In the next section \ref{section_numerical_applications}
are shown several numerical applications of the absolute 
seawater entropy, with: 
in subsection~\ref{subsection_SCIEX96} a study of a 
CTD (cast $43$) vertical profile observed during SCICEX'96,
with surface, Pacific, and Atlantic layers;  
in subsection~\ref{subsection_SCIEX97_transect} 
a study of two SCICEX'97 vertical transects; 
and in subsections~\ref{subsection_global_surf_entropy}
to \ref{subsection_global_surf_entropy_Western_Europe}
studies of the WOA23 analysed surface conditions  
at global and regional scales (Arctic Ocean, Bay of Bengal, 
Mediterranean, Black, and Caspian Seas).
It appears that the classic $t-S_{\rm A}$ ocean diagrams can be 
improved, with both the old isopycnic (iso-potential-density) 
lines and the new absolute iso-entropy lines, to show 
remarkable alignments of points along these two families 
of curves.

Finally, in Section~\ref{section_conclusion} I review the main results of the 
study and point out that turbulent processes may provide part of the explanation 
for these homogenisations of absolute seawater entropy, following the old 
Richardson's recommendations \citep[][]{Richardson_19a,Richardson_22}
that atmospheric and oceanic turbulence should apply to absolute entropy 
and not to temperature. 
\vspace*{-2mm}

 \section{Numerical applications}
\label{section_numerical_applications}
\vspace*{-2mm}

\begin{figure*}[tbp]
\centering
\includegraphics[width=0.336\linewidth,angle=0,clip=true]{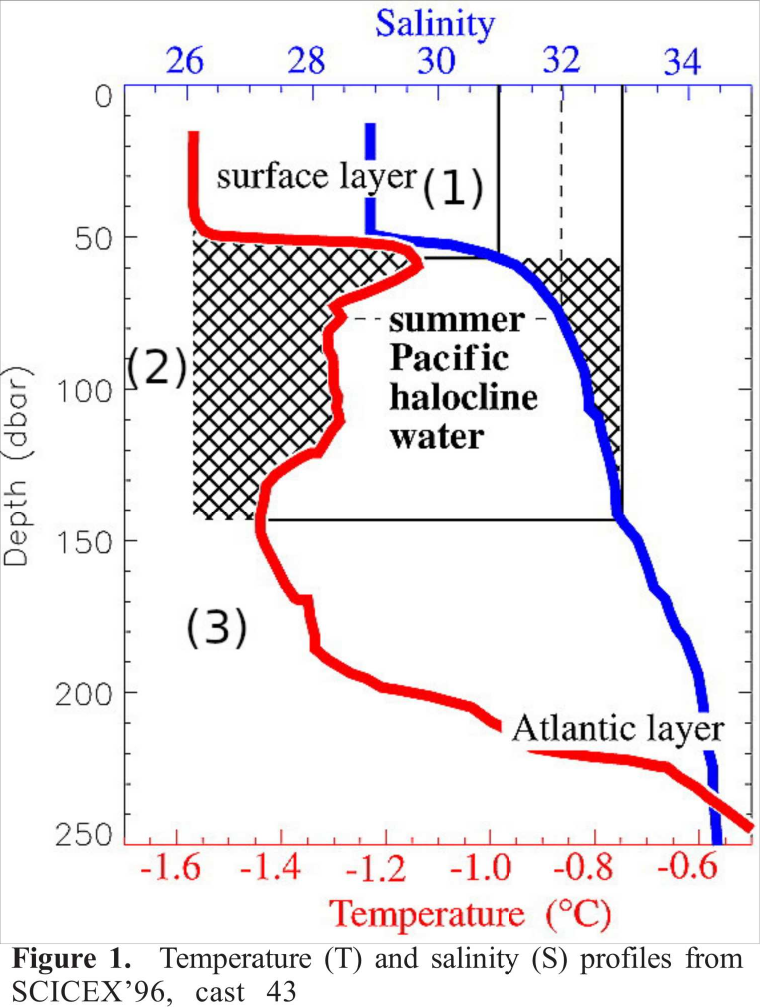}
\includegraphics[width=0.305\linewidth,angle=0,clip=true]{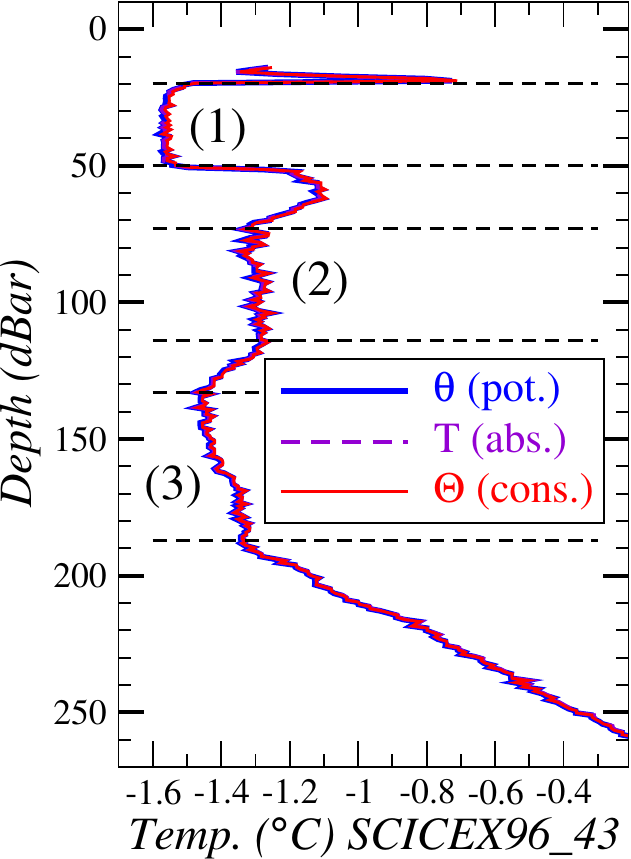}
\includegraphics[width=0.305\linewidth,angle=0,clip=true]{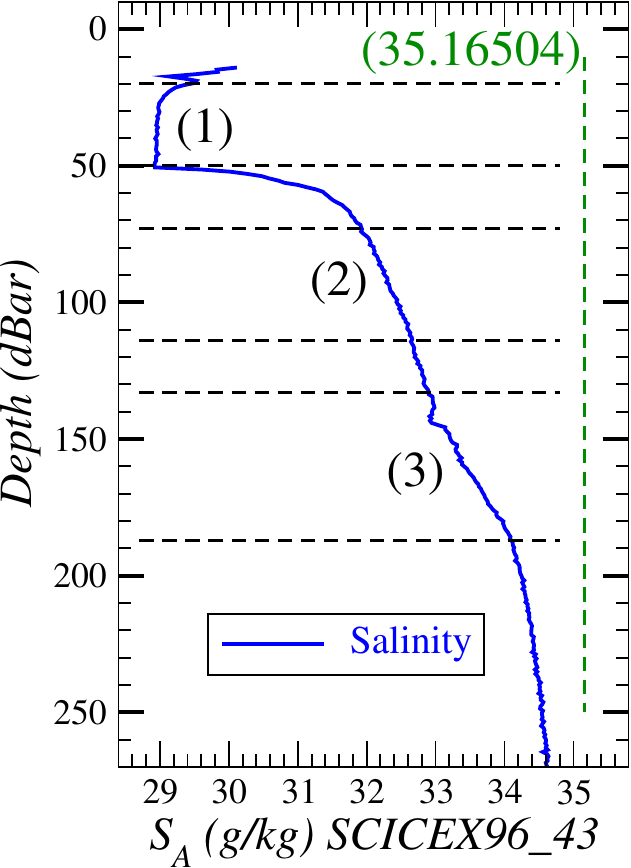}
\vspace*{2mm}\\
\includegraphics[width=0.41\linewidth,angle=0,clip=true]{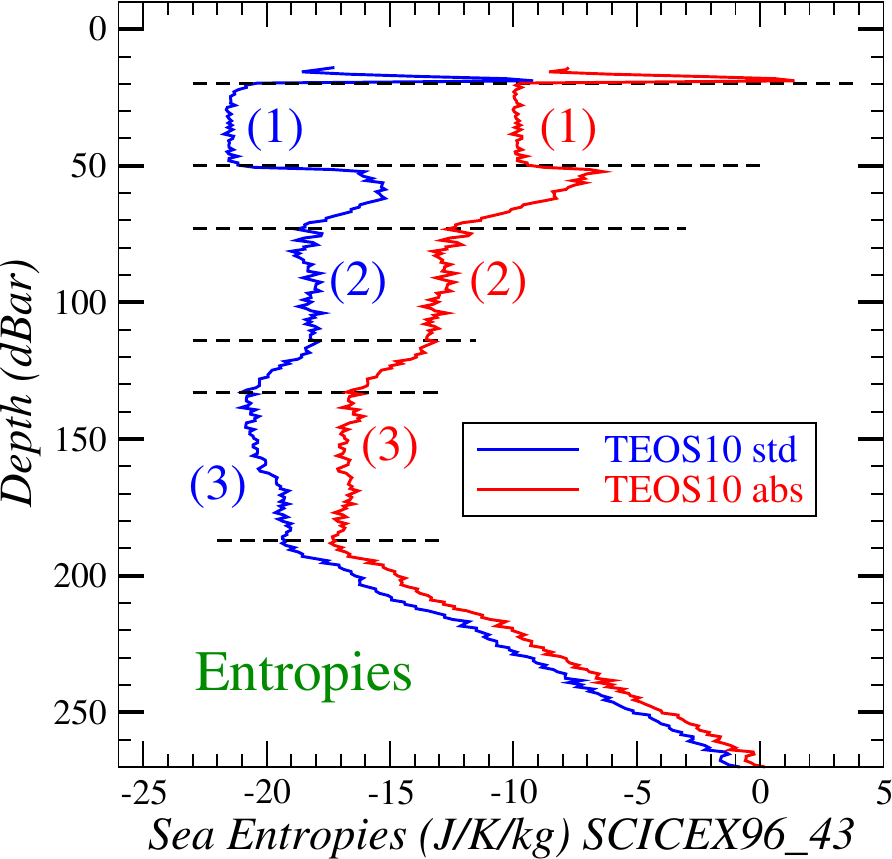} 
\includegraphics[width=0.45\linewidth,angle=0,clip=true]{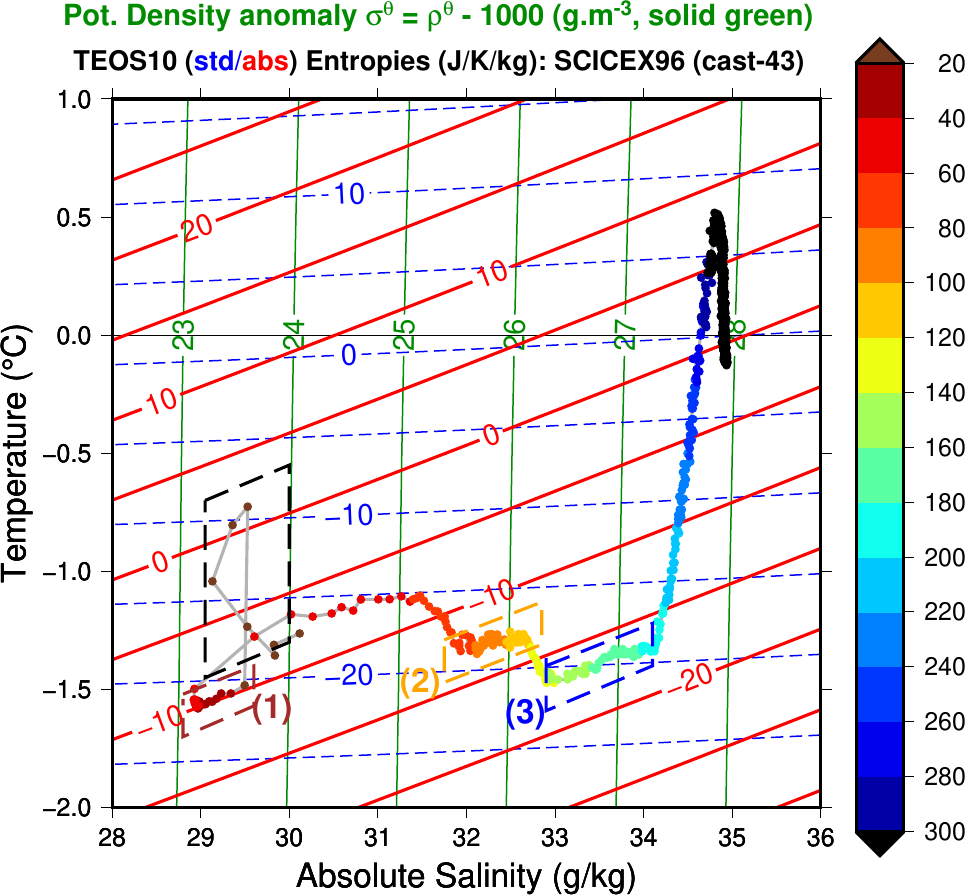}
\\
\vspace*{-2mm}
\caption{{ \it  \small
Top (left): an annotated version of Fig.~1 of \citet[][AGU grants]{Steele_al_JGR_2004} 
corresponding to a study of the low-resolution SCICEX'96 (cast 43) CTD vertical profiles. 
Top (middle and right): the corresponding full-resolution vertical profiles for the
potential temperature ($\theta$, solid blue), 
Celsius temperature ($T$,  dashed purple), 
conservative temperature ($\Theta$, solid red), 
and salinity (solid blue), 
with $S_{\rm SO}=35.165\,04$~g~kg${}^{-1}$ the TEOS10 standard salinity (green), 
for the high-resolution SCICEX'96 CTD (cast~43) vertical profiles. 
Bottom (left): the corresponding standard (solid blue) 
and absolute (solid red) versions of the seawater 
entropy, with the absolute mean value 
of the salinity increment $\Delta \eta_{\rm s}$ 
given by (\ref{Eq_etas_minus_etaw_value}).
Bottom (right): the same $t-S_{\rm A}$ diagram as in the Figs.~1 of Part~I, 
with the (almost horizontal) TEOS10's standard entropy (thin blue dashed lines),  
with the (more slantwise) TEOS10's absolute entropy (thick red solid lines),  
and with the plot of the SCICEX'96 (cast 43) vertical profile coloured from 
dark brown to dark blue for the depth from $14$~m to $300$~m,
and then in black up to the last $1004$~m depth. 
\label{Fig_TEOS10_SCICEX96_43}}\\
-----------------------------------------------------------------------------------------------------------------------------------------------
}
\end{figure*}

 \subsection{The SCICEX'96 experiment}
\label{subsection_SCIEX96}
\vspace*{-2mm}

To maximise the chances of observing significant impacts of salinity on absolute 
entropy profiles, it was necessary to find a well-documented example of an 
oceanic profile with both large variations in salinity and small variations 
in temperature.
The vertical profiles shown in the Figs.~\ref{Fig_TEOS10_SCICEX96_43}, 
already plotted by \citet[][his Fig.~1, p.~2]{Steele_al_JGR_2004}, 
can serve this purpose.

The aim of the paper by \cite{Steele_al_JGR_2004} was 
to study the circulation and interannual variability of 
summertime Pacific origin halocline waters in the 
Arctic Ocean, with the 
SCICEX'96$\,$\footnote{$\:$The general webpage of 
the ``SCICEX'' (Science Ice Exercise) 
is \url{https://nsidc.org/scicex/scicex-data}, 
with all the vertical CTD profiles available on the page  
\url{https://www.nodc.noaa.gov/archive/arc0021/0000568/1.1/data/0-data/}
(Submarine Arctic Science Program).} 
(cast 43) CTD vertical profiles typical of the Canadian Basin 
shown in Figure~\ref{Fig_TEOS10_SCICEX96_43}. 
This vertical profile is made of 
the ``summer Pacific halocline water'' layer (2), 
with a subsurface temperature maximum 
located in between the mixed ``surface layer'' (1) 
and the deeper ``Atlantic layer'' (3).

In the top panels 
are plotted:$\,$\footnote{$\:$The 
variables $\theta$ and $\Theta$ are computed 
via the subroutines   
\texttt{gsw\_pt0\_from\_t(S\_A,t,p)} and
\texttt{gsw\_ct\_from\_t(S\_A,t,p)} in the 
TEOS10-GSW software, from which the 
``potential enthalpy'' $h^0$ may be computed as 
$3991.868 \times \Theta$~J~kg${}^{\,-1}$
\citep[][]{McDougall_Potential_Enthalpy_2003,
McDougall_al_2023}.}
i)   the ``Celsius temperature'' $t$ (solid red); 
ii)  the ``potential temperature'' $\theta$ (dashed blue);    
iii) the seawater ``conservative temperature'' 
     $\Theta$ (dashed-dotted purple); and 
iv) the salinity (solid blue).
Top Figs.~\ref{Fig_TEOS10_SCICEX96_43} show that the high-resolution 
profiles indeed correspond (on average and below the large-temperature 
first $20$~m depth layer) to the mean profiles plotted 
in Fig.~1 of \citet[][]{Steele_al_JGR_2004}. 

The Celsius, potential, and conservative temperatures overlap 
almost exactly (up to less than $0.04$°C), with therefore 
very few differences in this Arctic case. 
This result is likely due to the small variations  
in temperature in this case, where $t$ remains between 
$-1.6$°C and $-0.4$°C.

Similarly, the bottom left part of Fig.~\ref{Fig_TEOS10_SCICEX96_43} 
shows that the structure of the vertical profile of the standard 
seawater entropy 
(in blue)$\,$\footnote{$\:$The standard TEOS10-GSW 
seawater entropy is computed via the subroutine 
\texttt{gsw\_entropy\_from\_t(S\_A,t,p)} in the 
TEOS10-GSW software.},
apart from changes in units, remains very close to that of 
the vertical temperature profile, with very little impact 
from strong variations in salinity. 
In particular, the values of the Pacific halocline layer (2)
are larger than both the surface layer (1) and deep Atlantic layer (3),
with values in (3) increasing with the depth for both the 
TEOS10 entropy and the Celsius temperature.

The fact that the TEOS10's version of the seawater entropy profiles 
closely resembles those of temperature can be understood as follows. 
Firstly, the impact of salinity $S_{\rm A}$ must be very low in the 
TEOS10's version, due to the hypothesis $(\eta_{\rm s0}-\eta_{\rm w0})=0$
and thus $\Delta \eta_{\rm s}=0$. 
Secondly, entropy must vary like the logarithm of absolute temperature 
via $C_p\:\ln(T/\overline{T})$, to within a constant. 
We can translate small variations in $T'=T-\overline{T}=t'$
by the second-order approximation $\ln(1+x)\approx x -x^2/2+...$, 
and thus the first-order result 
$\eta \approx (C_p/\overline{T})\:t' + cste$. 
This shows that the absolute seawater entropy  
is approximately proportional to $t' \approx 1$°C up to the 
quadratic term $-x^2/2$, and thus up to about  
$t'/(2\:\overline{T})\approx 0.5/273 \approx 0.2\,$\%. 

Differently, the vertical profile of absolute seawater entropy 
(in purple) shows different behaviors, with a decrease with the depth 
of the absolute entropy of the layer (1), then (2) and (3).
Note that the uncertainty of the impact of $\Delta \eta_{\rm s}$ is small (not shown), 
because the relative uncertainty of $-\,1880 \pm 17$ in 
(\ref{Eq_etas_minus_etaw_value}) represents only $1$~\% of the
difference between the blue and red curves in Fig.~\ref{Fig_TEOS10_SCICEX96_43}. 

In addition, the deep layer (3) becomes nearly isentropic with the 
absolute entropy formulation, due to compensation for temperature 
and salinity gradients that do not occur with the current TEOS10 
(arbitrary reference values) formulation.
The same is true at the top of the surface layer (1), where the 
increase in both temperature and salinity creates an extended 
absolute isentropic region including the $23$ and $19$~m layers, 
where the values of the standard TEOS10 entropy start to increase.

Another way to illustrate these features is the $t-S_{\rm A}$ diagram 
plotted in the bottom right part of Fig.~\ref{Fig_TEOS10_SCICEX96_43},
which shows in an interesting new way the three absolute isentropic regions 
(1), (2), and (3) that are clearly aligned along the new set 
of absolute isentropic red lines (see the three brown, 
orange, and blue dashed boxes), which cannot be due to chance.
The points are linked by a grey line in order to show how the 
upper levels are organised with a double loop (within the black 
dashed box) above and below the layer (1). 
The green isopycnic (potential-density anomaly) lines are almost vertical
and are not useful for studying the upper layers (1), (2), and (3),
even though the deep-layer points ($z<190$~m depth) roughly converge 
toward the isopycnic line $\sigma^{\theta} \approx 27.8$~g~m${}^{-3}$.

There may therefore be a double set of physical constraints 
illustrated by the two sets of curves in the $t-S_{\rm A}$ diagrams:
following either the absolute-isentropic red lines or the isopycnic 
lines (iso-pot.-density, in green), as confirmed by the 
other similar results shown in the next subsections. 

The standard TEOS10 entropy (dashed thin blue) lines are almost 
isothermal lines for temperatures between $-2$°C and $+1$°C.  
The black points and dashed box represent the very surface 
atypical conditions (between $14$ and $19$~m depth). 
The graphical study of vertical ocean profiles could therefore 
be revisited thanks to this kind of classical $t-S_{\rm A}$ diagram, 
but only if including the new set of red lines to reveal the absolute 
isentropic regions. 

The absolute isentropic conditions of these three layers (1), (2)
and (3) can also be understood from (\ref{Eq_etas_minus_etaw}),
(\ref{Eq_etas_minus_etaw_value}), and with the first-order 
result $\eta \approx (C_p/\overline{T})\:t' + cste$. 
It is indeed possible to provide a first-order condition
$(\delta \eta=0)$ for the respective impacts of joint variations 
in temperature $(\delta t)$ and salinity $(\delta S_{\rm A})$,
leading to
\begin{align}
  \frac{C_p}{\overline{T}} \times \delta t
      & \: \leftrightarrow \;  
  -\,\frac{(\eta_{\rm s0}-\eta_{\rm w0})}{1000} \times \delta S^A_{\rm A}
\nonumber \; , \\
   14 \times \delta t 
    & \: = \: 1.88 \: ( 7.4 \times \delta t )
      \: \leftrightarrow \; 1.88 \: (\delta S_{\rm A})
\label{Eq_detas_dt_dSA} \; , \\
  \delta \eta 
    & \: \approx \: 1.88 \: 
      ( 7.4 \times \delta t - \: \delta S_{\rm A} )
\label{Eq_detas_dt_dSA_bis} \; ,
\end{align}
where $C_p \approx 4218$~J~K${}^{\,-1}$~kg${}^{\,-1}$,  
$\overline{T} \approx 300$~K, $t'$ in °C and 
$S_{\rm A}$ in g~kg~${}^{\,-1}$.
An application to the deep layer (3) for the SCICEX'96 
(cast 43) CTD vertical profile corresponds to 
$\delta t \approx 0.17$~K and 
$\delta S_{\rm A} \approx 1.2$~g~kg~${}^{\,-1}$,
which from (\ref{Eq_detas_dt_dSA}) indeed corresponds to
$ 7.4 \times 0.17 = 1.26 \approx 1.2$ and to 
nearly isentropic conditions.
Note that the draft value $14$ in (\ref{Eq_detas_dt_dSA})
agrees with the linear term in $y$ in 
Table~1 in Part~I: 
$[\:24715.571866078/(40 \times 40)\:] \times t
\approx 15.4 \times t$.
Note also that these relationships (\ref{Eq_detas_dt_dSA}) and (\ref{Eq_detas_dt_dSA_bis}) 
are only valid for temperatures close to $0$°C, with $|y|=|t|/40 < 0.05$ small and 
thus with all the high-order terms depending on $y^2$, $y^3$, ... 
duly neglected.

The presence of the surface fresh water with low salinity 
(here $<30$~g~kg${}^{\,-1}$) results in a large increase 
in absolute entropy ($+11.5$~J~K${}^{-1}$~kg${}^{-1}$) for this SCICEX'96 
cast 43 profile due to $\Delta \eta_{\rm s}$, but with no impact on the usual 
TEOS10's version of entropy.
However, these impacts cannot be arbitrary because, otherwise, there would be no 
point in measuring the entropy of the ocean. 
Indeed, the reference value $(\eta_{\rm s0}-\eta_{\rm w0})$ has clear impacts on   
  i) the difference in entropy between two points due to 
  $(\Delta \eta_{\rm s})_2 - (\Delta \eta_{\rm w})_1$ and thus 
  $(\eta_{\rm s0}-\eta_{\rm w0})/1000 \times [\:(S_{\rm A})_1-(S_{\rm A})_2)\:]$;  
  ii) the gradient in entropy due to
    $(\eta_{\rm s0}-\eta_{\rm w0})/1000 \times \nabla S_{\rm A}$; and  
  iii) the temporal evolution in entropy due to
    $(\eta_{\rm s0}-\eta_{\rm w0})/1000 \times [\:dS_{\rm A}/dt\:]$. 
It is therefore not possible to consider that $(\eta_{\rm s0}-\eta_{\rm w0})$ 
could have no impact on the computation of a thermodynamic state 
function such as entropy; otherwise, this would be equivalent to 
denying this state function any physical meaning.

For instance, the second law of thermodynamics 
may imply, for suitable conditions, that the 
entropy should increase monotonically until 
it reaches its maximum at the state of
thermodynamic equilibrium.
But the search for such a ``maximum entropy state'' 
would be meaningless if one could modify at will 
the values of the seawater entropy, with, for instance, 
$\Delta \eta_{\rm s}=0$ presently 
assumed in the TEOS10 version and in 
all other present studies, with detrimental 
impacts in the entropy profiles in the 
Figs.~\ref{Fig_TEOS10_SCICEX96_43}.
Similarly, the isentropic aspect of the deep layer (3) 
only appears with the absolute version of entropy, which 
is the only way of revealing this aspect, which acquires 
an obvious physical meaning and which cannot be modified 
by arbitrary choices of $\eta_{\rm s0}$ and $\eta_{\rm w0}$,  
as presently made in TEOS10 and contrary to the third 
law of thermodynamics.

\begin{figure*}[hbt]
\centering
\includegraphics[width=0.8\linewidth,angle=0,clip=true]{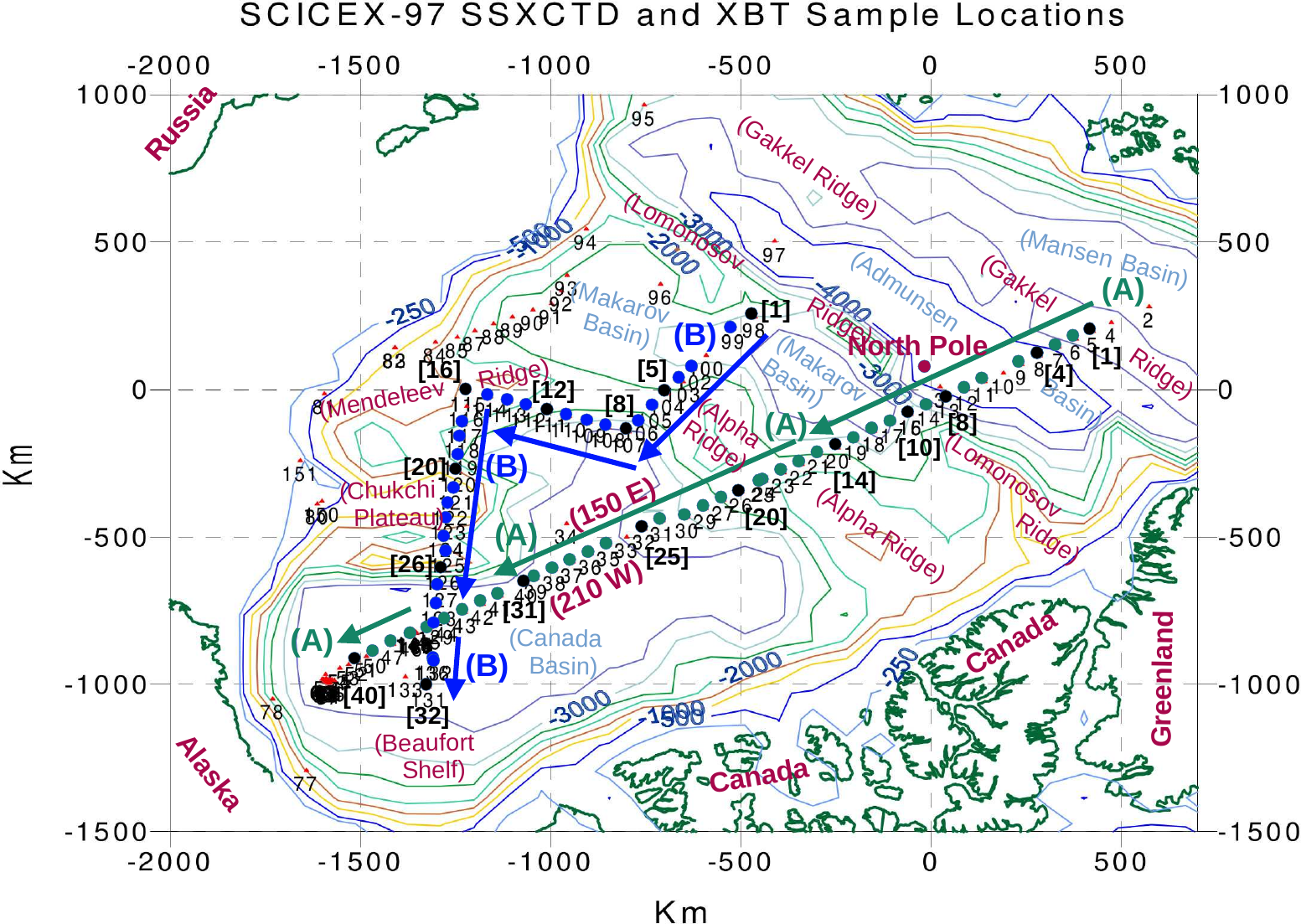}\\
\includegraphics[width=0.49\linewidth,angle=0,clip=true]{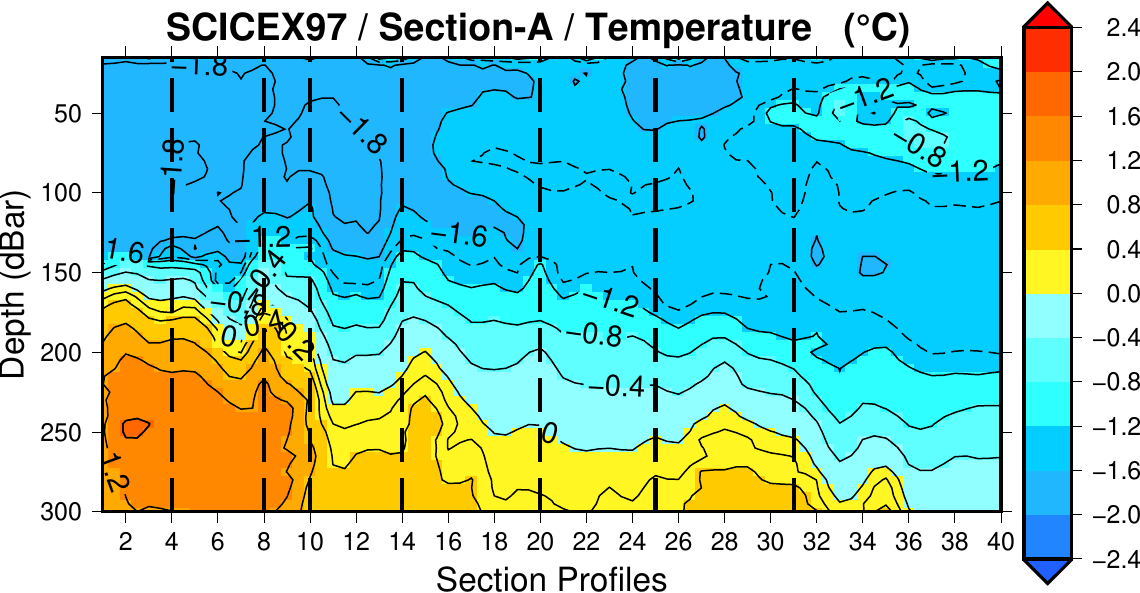}
\includegraphics[width=0.49\linewidth,angle=0,clip=true]{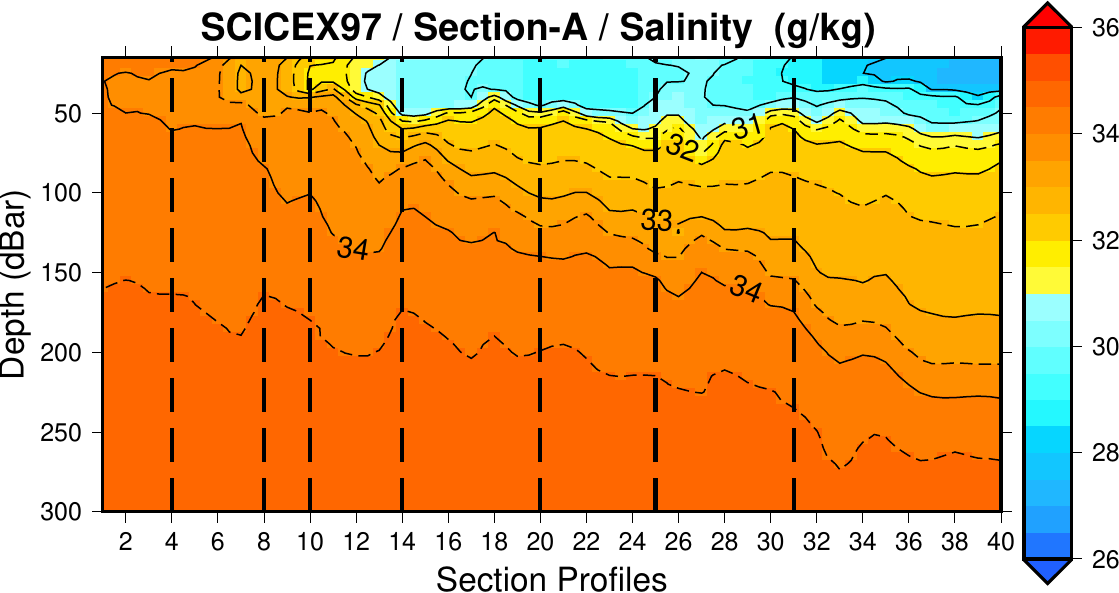}
\vspace*{1mm} \\
\includegraphics[width=0.49\linewidth,angle=0,clip=true]{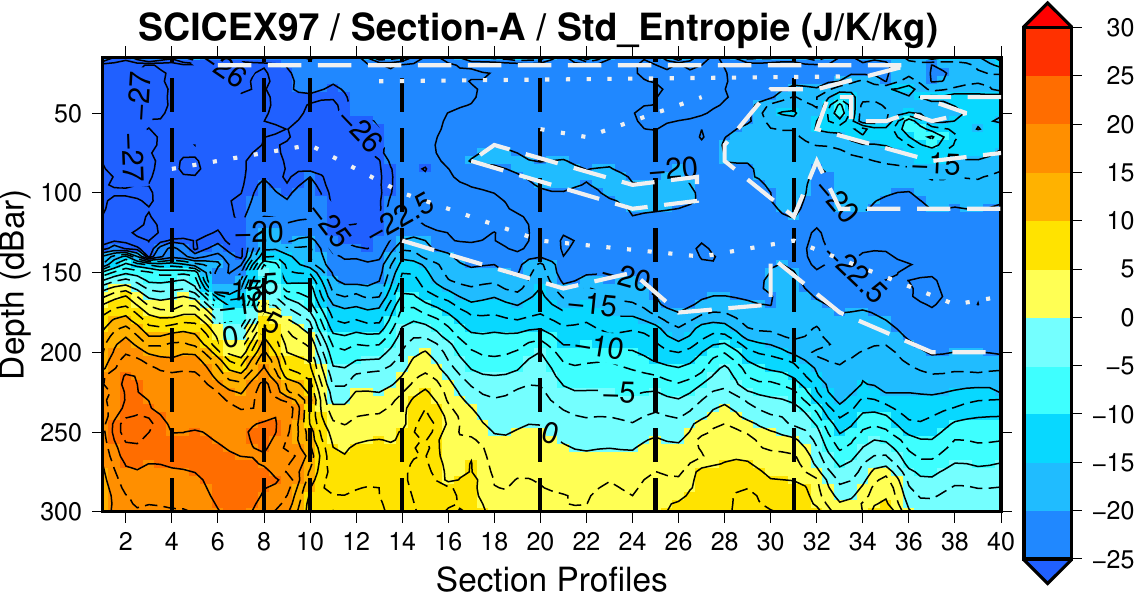}
\includegraphics[width=0.49\linewidth,angle=0,clip=true]{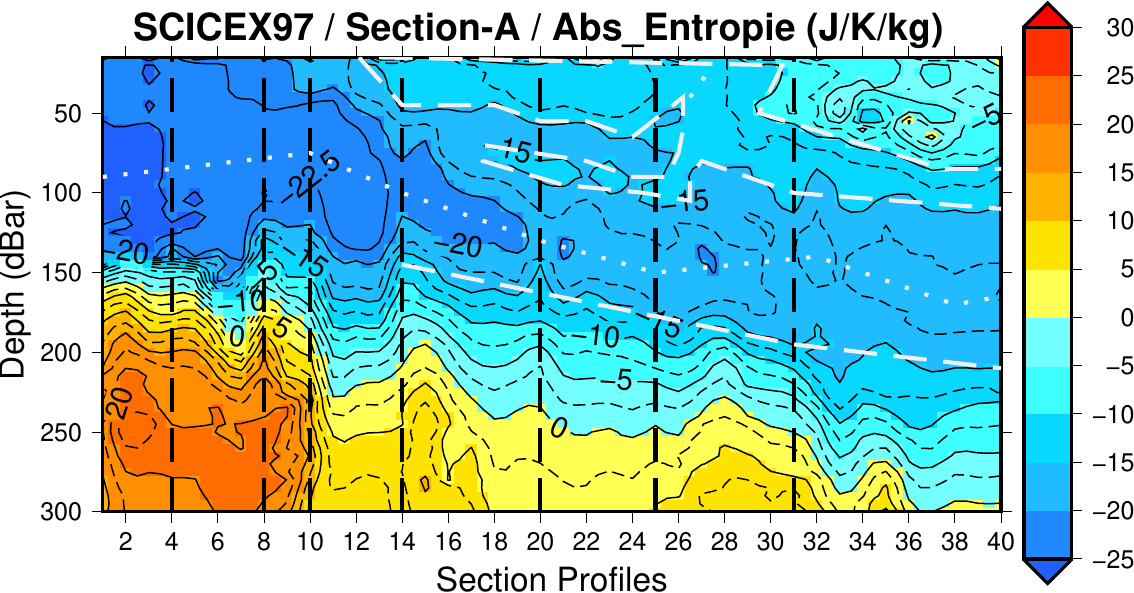}\\
\caption{{\it\small On the top: the annotated SCICEX-97 Sample Locations map 
(green points for the Transect-A and blue points for the Transect-B). 
Then the four Transect-A figures (from $15$~m to $300$~m depth) 
for the temperature (°C), salinity (g~kg${}^{\,-1}$), standard (TEOS10), 
and absolute (TEOS10+third-law) seawater entropies (J~K${}^{\,-1}$~kg${}^{\,-1}$). 
\label{Fig_SCICEX97_transect_A}}\\
-----------------------------------------------------------------------------------------------------------------------------------------------
}
\end{figure*}
\clearpage

\begin{figure*}[hbt]
\centering
\includegraphics[width=0.49\linewidth,angle=0,clip=true]{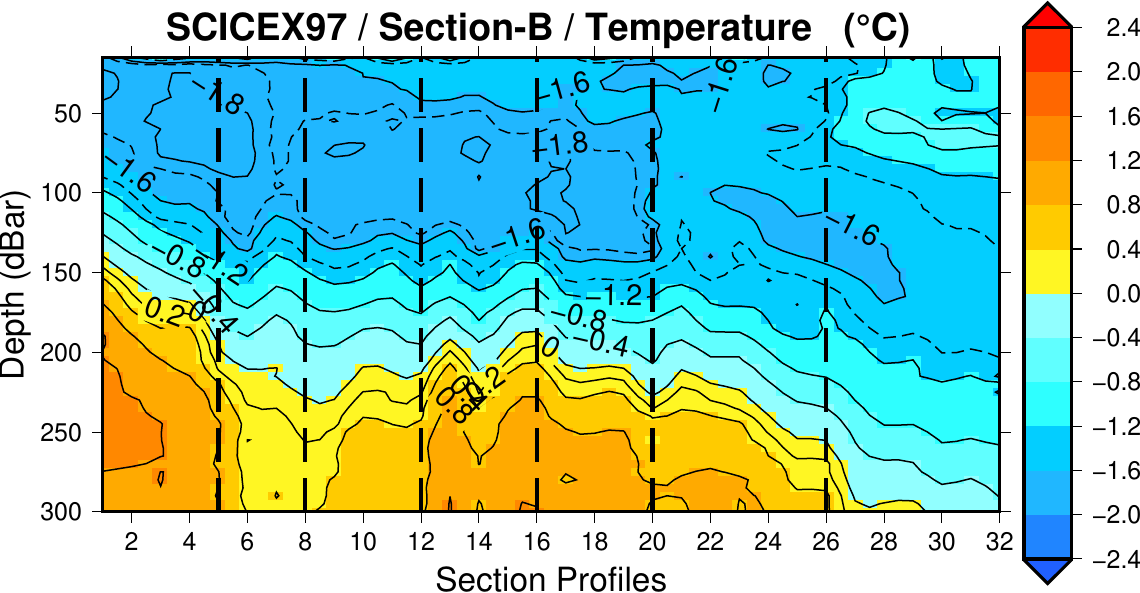}
\includegraphics[width=0.49\linewidth,angle=0,clip=true]{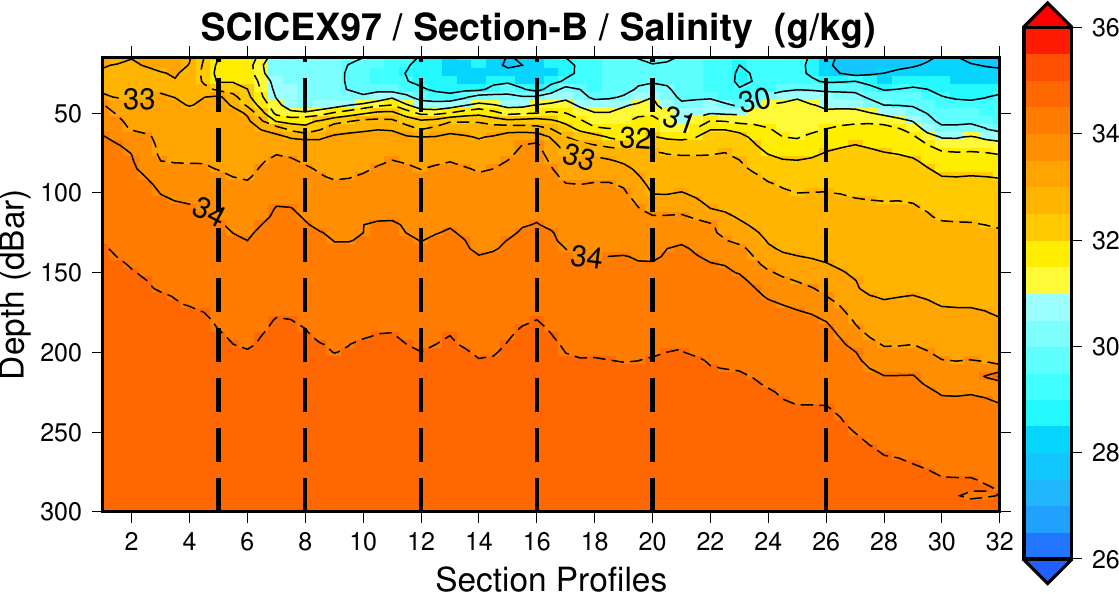}\\
\includegraphics[width=0.49\linewidth,angle=0,clip=true]{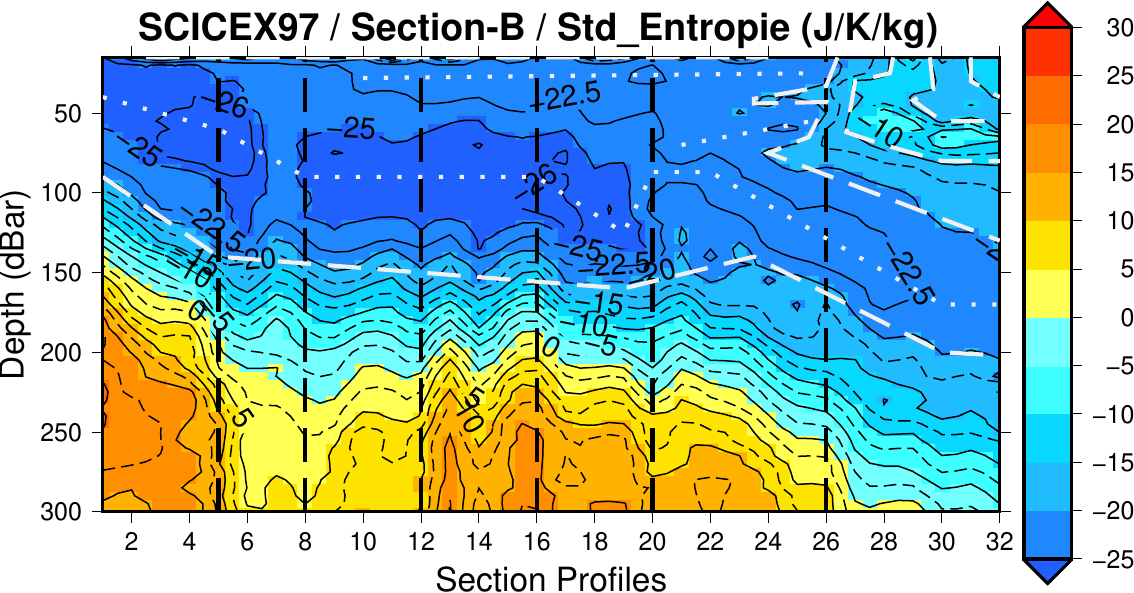}
\includegraphics[width=0.49\linewidth,angle=0,clip=true]{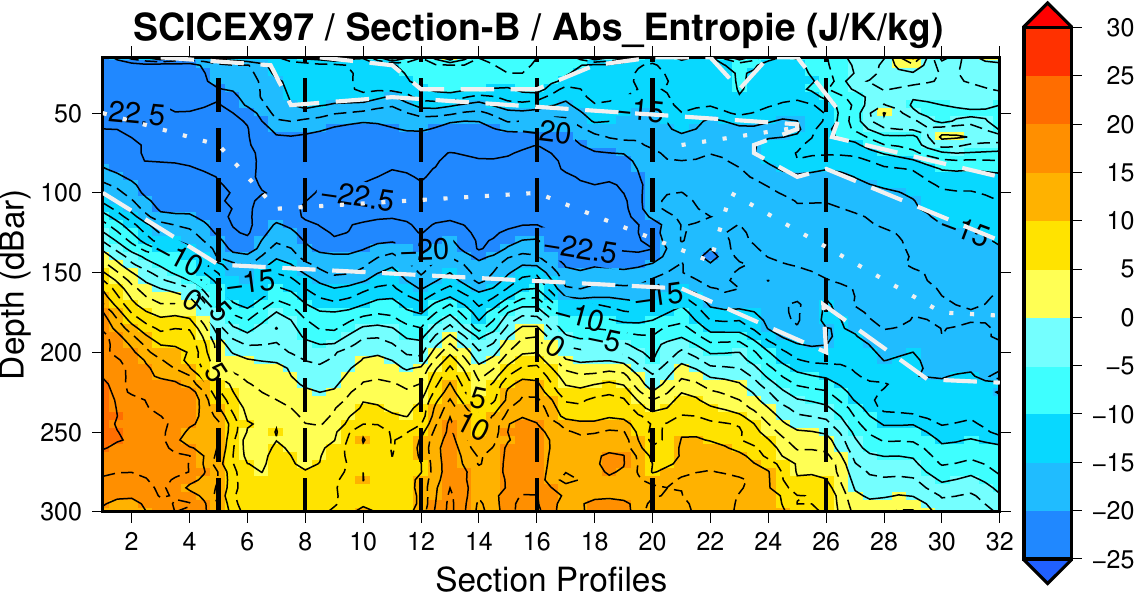}\\
\caption{{\it\small Same as for Figs.~\ref{Fig_SCICEX97_transect_A}, but for the Transect-B figures.
\label{Fig_SCICEX97_transect_B}}\\
-----------------------------------------------------------------------------------------------------------------------------------------------
}
\end{figure*}

 \subsection{The SCICEX'97 transects}
\label{subsection_SCIEX97_transect}
\vspace*{-2mm}

The properties highlighted on the sole vertical profiles of SCICEX'96 showing 
the interest of defining the seawater entropy in an absolute way 
(as suggested by thermodynamics) need to be confirmed on other examples. 
To do this, before studying even more global properties in the other sections, 
it is possible to study the SCICEX'97 transects, which are a kind of spatial 
extension of the SCICEX'96 profile.

I have retained $40$ of the CTD0xx profiles to make the 
transect~A shown in the Figs.~\ref{Fig_SCICEX97_transect_A}, 
with xx from $05$ to $52$ (except the unusual noisy $15$, $
28$, $33$, $34$, $48$, $49$, and $51$ 
CTD)\footnote{$\:$The transect~A
(green points in the SCICEX-97 Sample Locations map)
was roughly located along 
the longitude $150$~E ($210$~W) and extended from 
the Gakkel Ridge (points $[1]$ to $[3]$ for CTD005 to CTD007); 
the Admunsen Basin (points $[4]$ to $[7]$); 
the Lomonosov Ridge (close to the North Pole, points $[8]$ to $[10]$); 
the Makarov Basin (points $[11]$ to $[13]$); 
the Alpha Ridge (points $[14]$ to $[20]$); 
the following Shelf (points $[21]$ to $[25]$); 
and finally the Canada Basin (points $[25]$ to $[40]$), 
passing close to the Chukchi Plateau (point $[31]$).}. 
I have retained $32$ of the CTD0xxx profiles to make the 
transect~B shown in the Figs.~\ref{Fig_SCICEX97_transect_B}, 
with xxx from $098$ to $129$, $132$, $136$, and $131$ (except the 
unusual noisy $100$, $101$, $116$, $130$, and $133$ to $135$  
CTD)\footnote{$\:$The transect~B 
(blue points in the SCICEX-97 Sample Locations map)
extended from 
the Makarov Basin (points $[1]$ to $[4]$ for CTD098 to CTD102); 
the Alpha Ridge (points $[5]$ to $[7]$);
the Canada Basin Shelf (points $[8]$ to $[11]$); 
the Mendeleev Ridge (points $[12]$ to $[16]$); 
the Canada Basin Shelf (points $[17]$ to $[19]$); 
the Chukchi Plateau (points $[20]$ to $[25]$);
and finally the Canada Basin (points $[26]$ to $[32]$).}. 

The two transects, A and B, exhibit almost the same properties, 
with a roughly north-south variation of the points and with 
an absolute (thermodynamic) version of the seawater entropy 
more interesting and more coherent spatially than the standard 
(TEOS10) arbitrary version.

Indeed, if the subsurface layer (between $15$ and $150$-$200$~m) 
is partly associated with the minimum of temperature (from $-1.2$ to $-1.85$°C), 
it should also be modulated by the vertical gradient of salinity 
(from $28$ to $34$~g~kg${}^{\,-1}$).
But this is only the case due to the impact of the salinity increment 
$\Delta \eta_{\rm s}$ on the absolute seawater entropy, which is to create 
a more coherent and unique subsurface layer with minimum values 
from $-15$ to $-25$~J~K${}^{\,-1}$~kg${}^{\,-1}$.
Differently, the standard version shows a more uneven shape with 
a second layer (at about $30$~m depth for all latitudes) 
largely influenced by the associated cold temperature tongue, 
without the thermodynamic impact of $\Delta \eta_{\rm s}$ 
that creates the unique layer. 
There is also a clear deepening with decreasing latitudes for 
the absolute version (see the white dashed and dotted annotating lines).

Another interesting feature is the more coherent fresh surface layers 
with low salinity, with associated larger absolute values 
of the seawater entropy (light-blue and yellow colours, 
from $-5$ to $+5$~J~K${}^{\,-1}$~kg${}^{\,-1}$) 
and the greater values at the surface, 
whereas they are in a subsurface position for the standard 
(TEOS10) seawater entropy values (at about $60$~m depth).
We can see here the same impact already observed with the 
SCICEX'96 profile and layer (1), which has a higher absolute 
entropy than the other subsurface layers (2) and (3), 
whereas the opposite is true for the first layer (1), 
which has the lowest standard TEOS10 entropy.

All these differences confirm that the arbitrary definitions leading 
to the standard value of the TEOS10 entropy lead to significant 
differences in the spatial variations (in latitudes and depths) 
of the seawater entropy in comparison with its thermodynamic 
absolute definition. 
This could provide an incentive to plot and enrich future studies 
of this new thermodynamic variable, including the salinity 
increment term $\Delta \eta_{\rm s}$.

 \subsection{A study of global surface ocean entropy}     
\label{subsection_global_surf_entropy}
\vspace*{-2mm}

We cannot deduce from the study of the previous few examples that absolute 
entropy is of interest elsewhere than in the polar regions, or even that 
it is of interest everywhere in these polar regions. 
We need to carry out other, more systematic, and therefore more global, 
studies.

\begin{figure*}[hbt]
\centering
\includegraphics[width=0.49\linewidth,angle=0,clip=true]{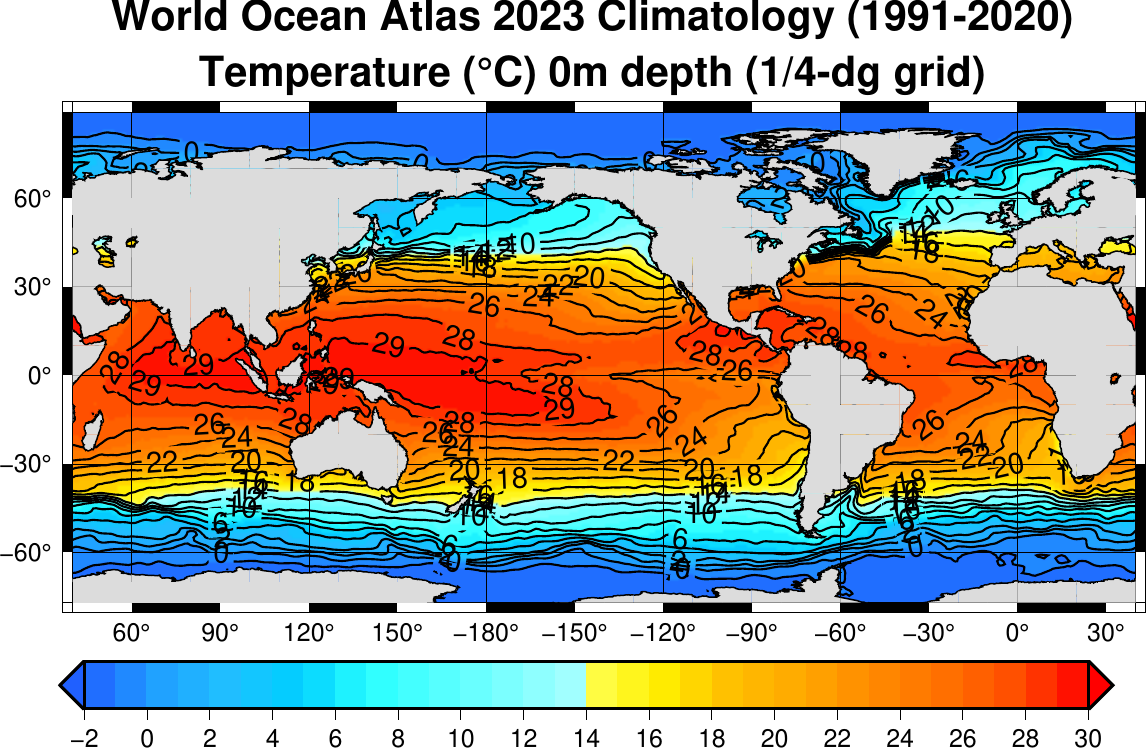}
\includegraphics[width=0.49\linewidth,angle=0,clip=true]{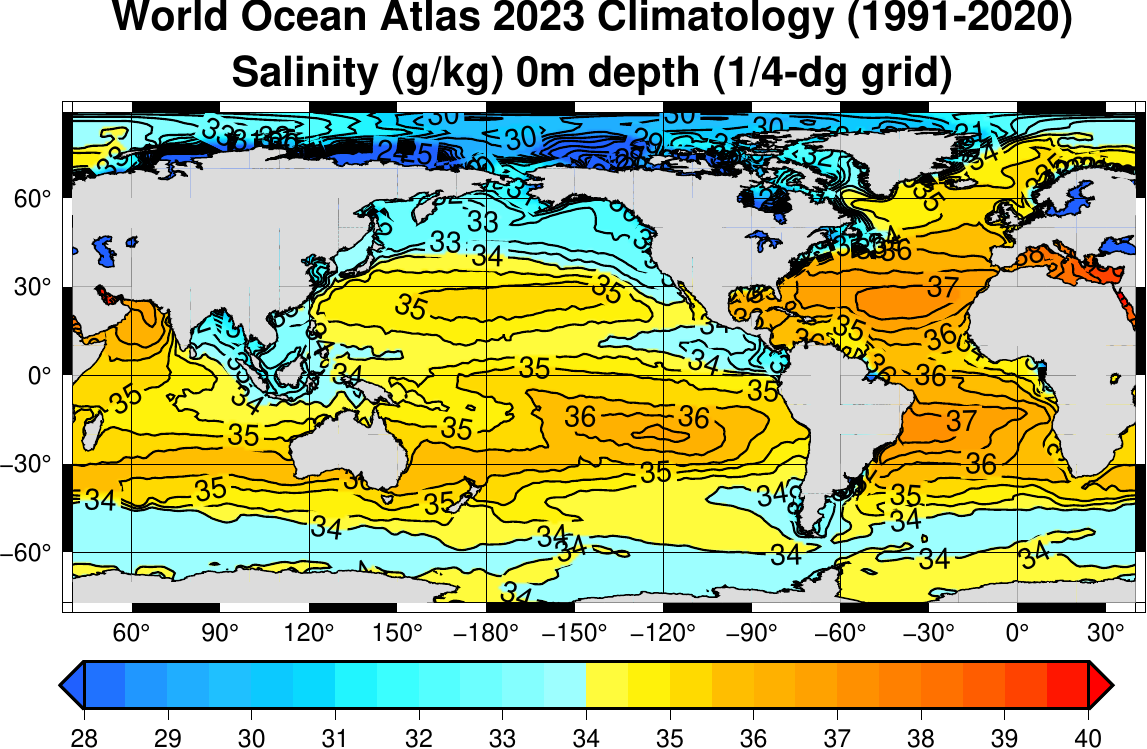}
\vspace*{2mm}\\
\includegraphics[width=0.49\linewidth,angle=0,clip=true]{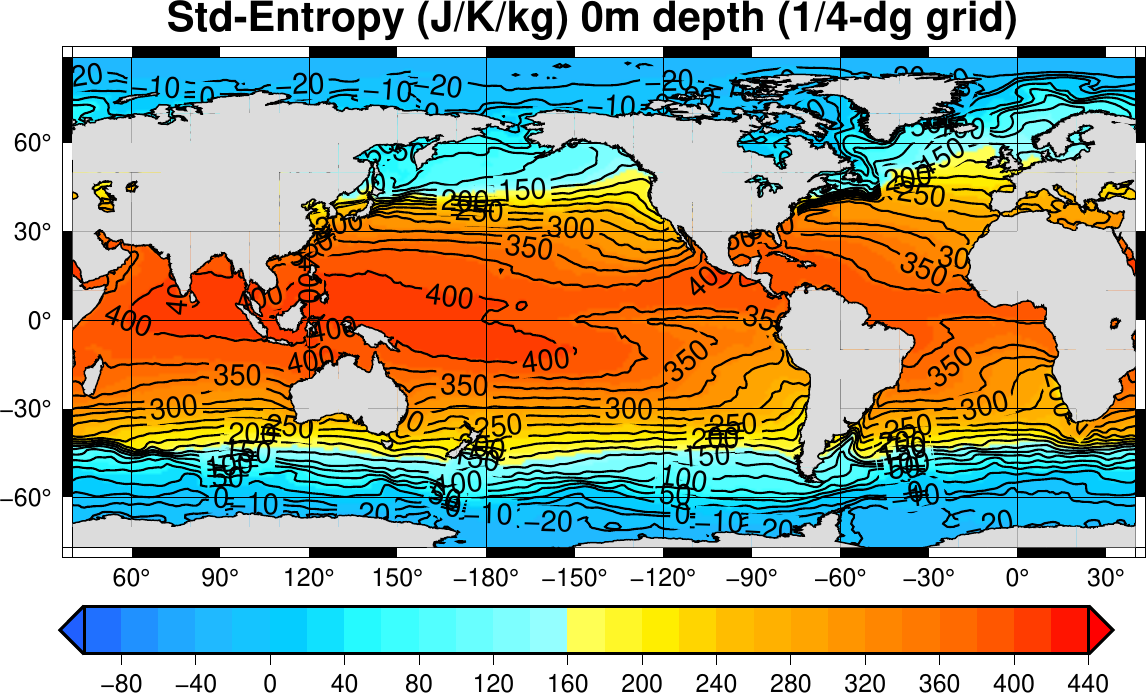}
\includegraphics[width=0.49\linewidth,angle=0,clip=true]{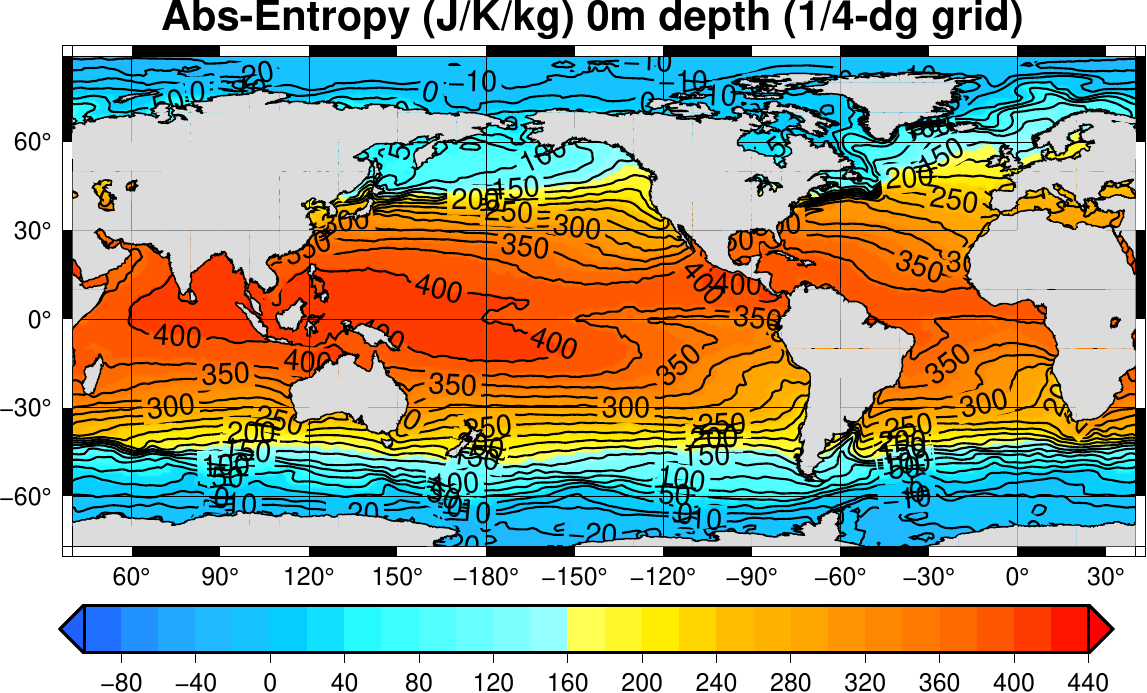}
\vspace*{2mm}\\
\includegraphics[width=0.75\linewidth,angle=0,clip=true]{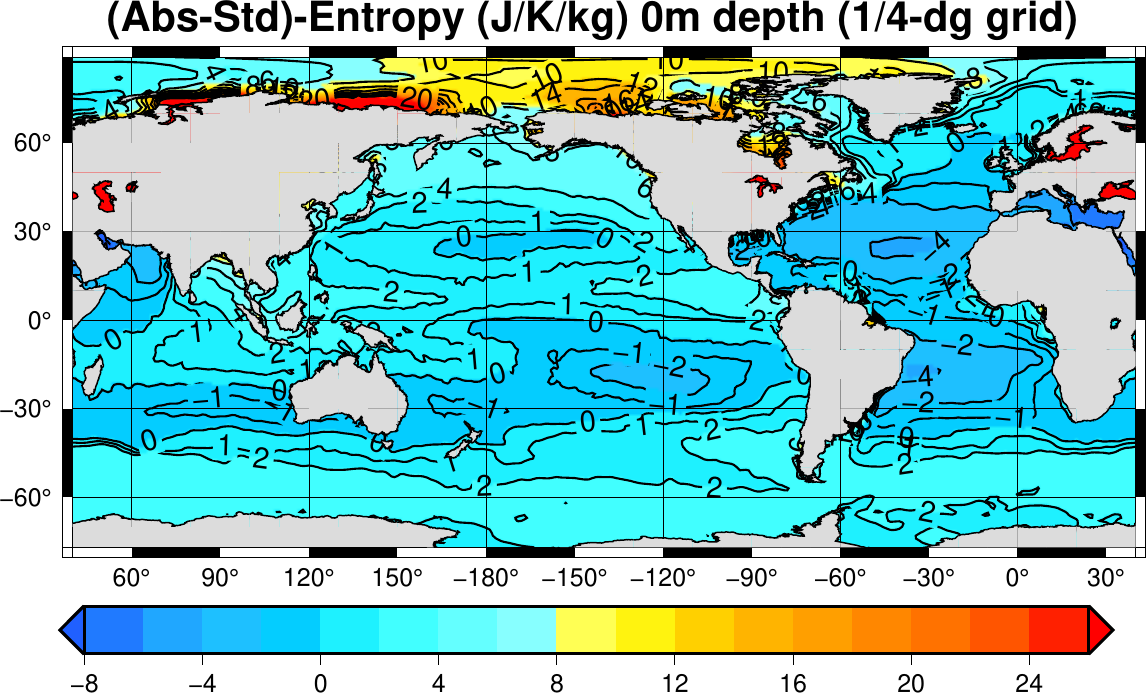}
\\
\caption{{\it\small The 1991-2020 objectively analyzed annual surface means 
for the temperature, salinity, standard (Std) seawater entropy, 
absolute (Abs) seawater entropy,   
and difference (Abs$\,-\,$Std) in seawater entropy, 
computed from the World Ocean Atlas 2023 
climatology (WOA23, release February 2024, quarter-degree grid). 
Documentation and data are available at 
\url{https://www.ncei.noaa.gov/products/world-ocean-atlas}.
\label{Fig_WOA23_global}}\\
-----------------------------------------------------------------------------------------------------------------------------------------------
}
\end{figure*}

At first sight, the global study shown in Figs.~\ref{Fig_WOA23_global} 
suggests that the surface mean standard (TEOS10) seawater entropy 
is almost everywhere similar to the absolute (third-law) version 
computed with the 1991-2020 annual means from the World Ocean Atlas 2023 
(WOA23, quarter-degree resolution).
This result is a priori somewhat disappointing, in the same ways as the  
potential and conservative temperatures almost overlap the absolute temperature 
for the SCICEX'96 vertical profile shown in Fig.~\ref{Fig_TEOS10_SCICEX96_43}. 

However, several differences can be guessed at the regional scale, such as 
those shown in Figs.~\ref{Fig_WOA23_regional_Arctic}, 
\ref{Fig_WOA23_regional_IndoAsie}, and \ref{Fig_WOA23_regional_Mediter} 
and mainly corresponding to the regions with either large gradients in salinity  
or small ($<32$~g~kg${}^{\,-1}$) or large ($>37$~g~kg${}^{\,-1}$) values 
of the salinity itself, as shown in many places in the 
last panel of Fig.~\ref{Fig_WOA23_global} for the difference (Abs$\,-\,$Std) 
in seawater entropy (dark blue negative and yellow, orange, and red positive values).
Among these regions are the Arctic Ocean, the Bay of Bengal,  
and the Northeast Atlantic and Mediterranean Seas.

The noisy appearance of the contours of certain (global) fields plotted 
with the GMT graphic tool is similar to the noise clearly existing in 
the corresponding graphs (not shown) plotted directly from the WOA23 
site and with the same quarter-degree resolution.

Note that the influence of non-linearities in the TEOS10 entropy 
formula is small (not shown), with  similar values for the 
entropy $\eta(\overline{t},\overline{S_{\rm A}})$ 
computed from the annual mean values 
$\overline{t}$ and $\overline{S_{\rm A}}$ for temperature and salinity, 
on the one hand, and for the mean entropy averaged from the $12$ monthly 
mean values $\overline{\eta(t,S_{\rm A})}$, 
on the other hand.

 \subsection{A study of surface entropy: Arctic Ocean}
\label{subsection_global_surf_entropy_Arctic =_ocean}
\vspace*{-2mm}

\begin{figure*}[hbt]
\centering
\includegraphics[width=0.49\linewidth,angle=0,clip=true]{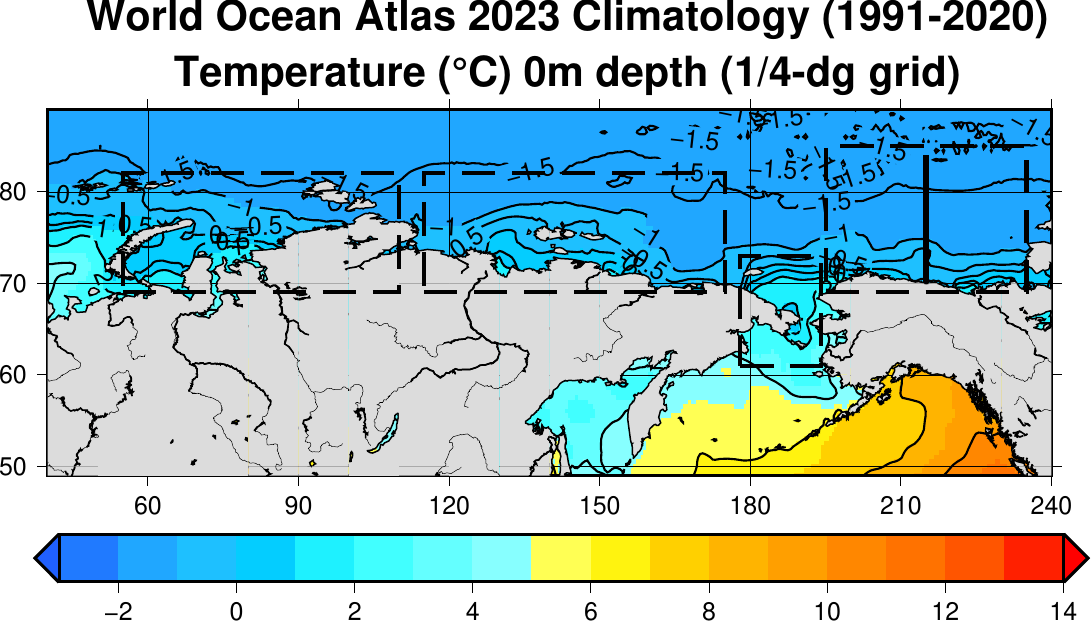}
\includegraphics[width=0.49\linewidth,angle=0,clip=true]{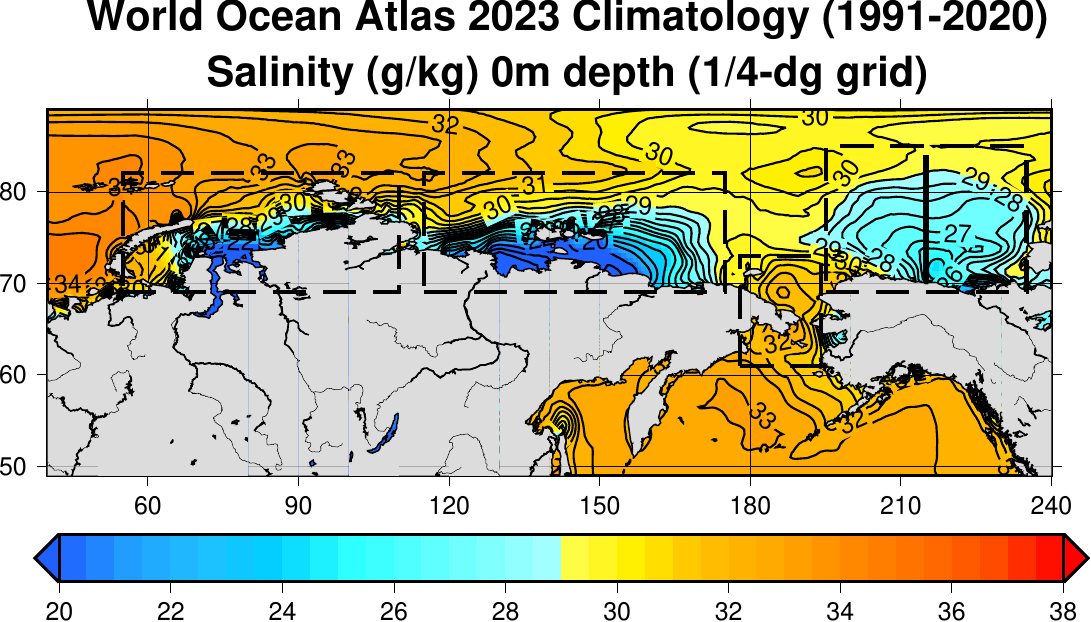}
\vspace*{2mm}\\
\includegraphics[width=0.49\linewidth,angle=0,clip=true]{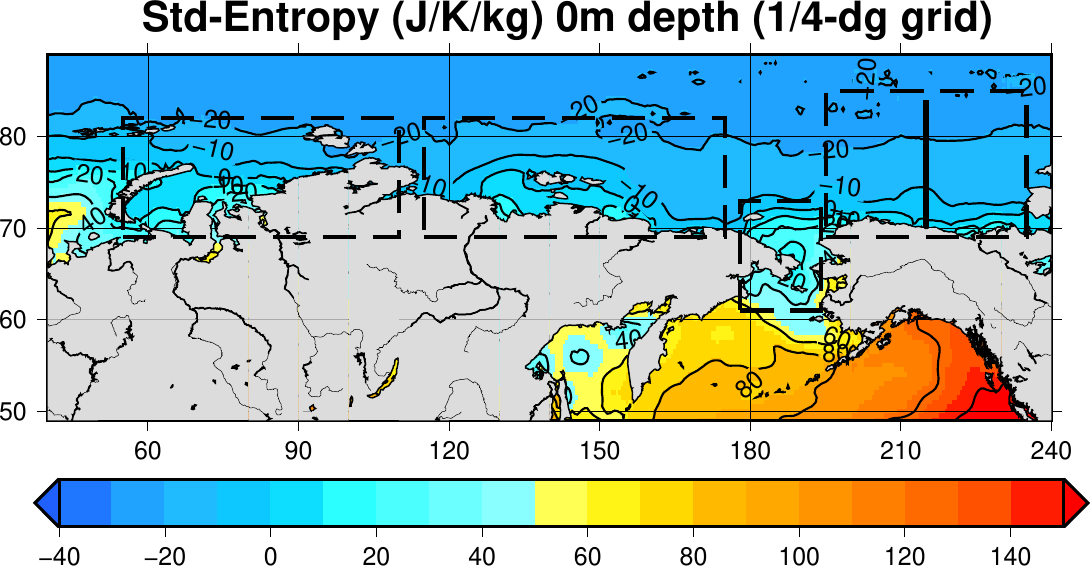}
\includegraphics[width=0.49\linewidth,angle=0,clip=true]{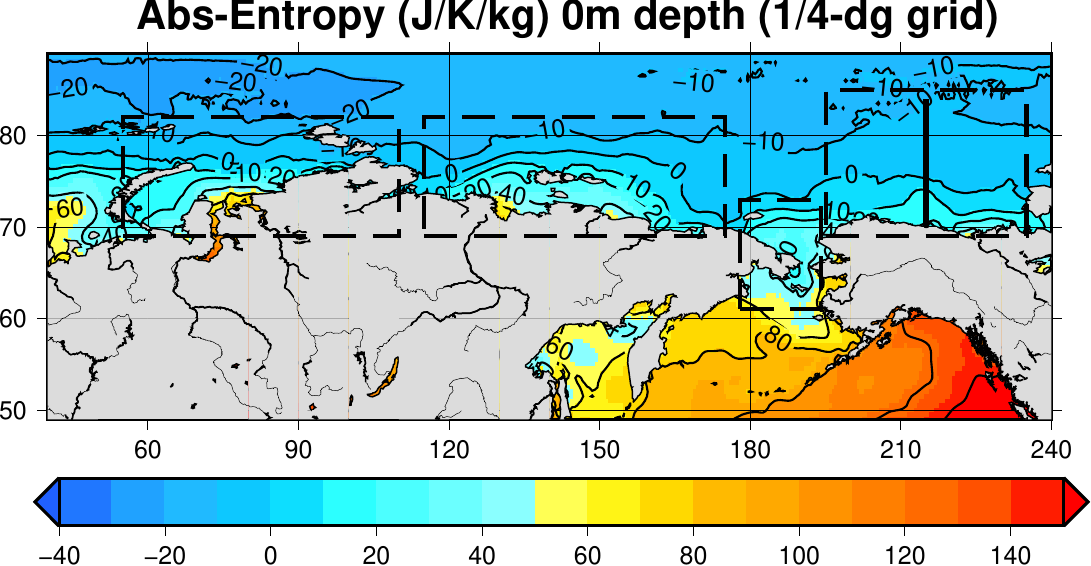}
\vspace*{2mm}\\
\includegraphics[width=0.49\linewidth,angle=0,clip=true]{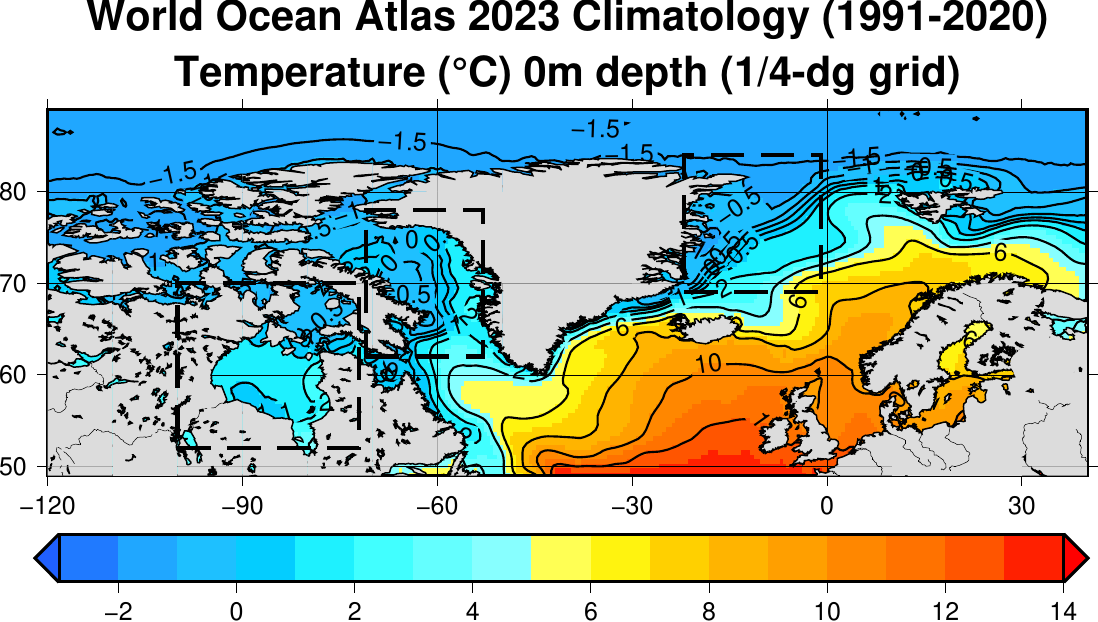}
\includegraphics[width=0.49\linewidth,angle=0,clip=true]{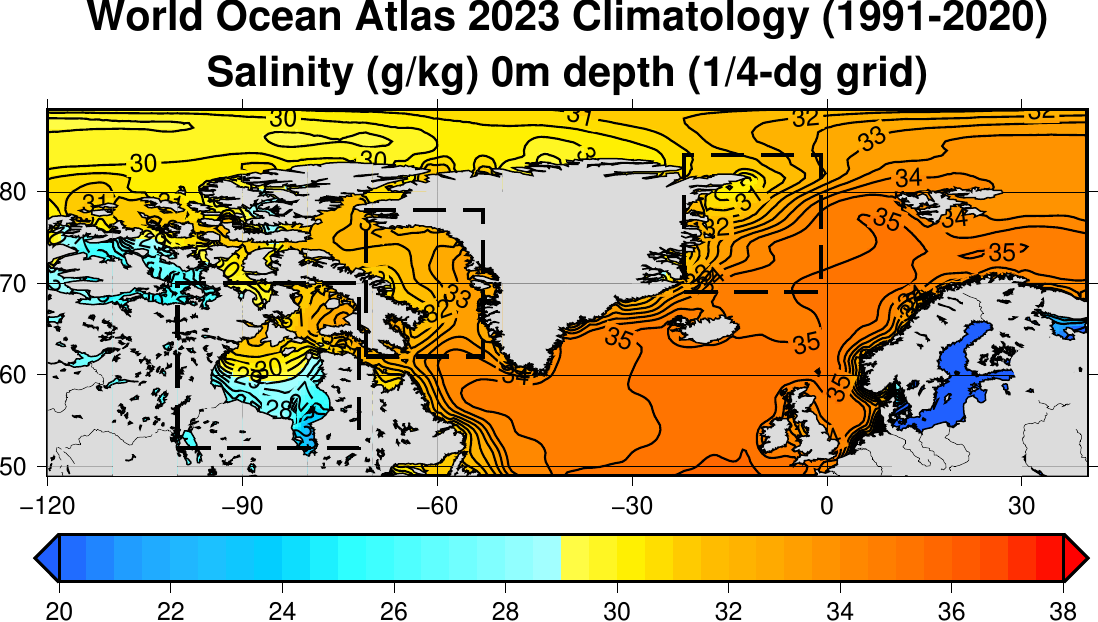}
\vspace*{2mm}\\
\includegraphics[width=0.49\linewidth,angle=0,clip=true]{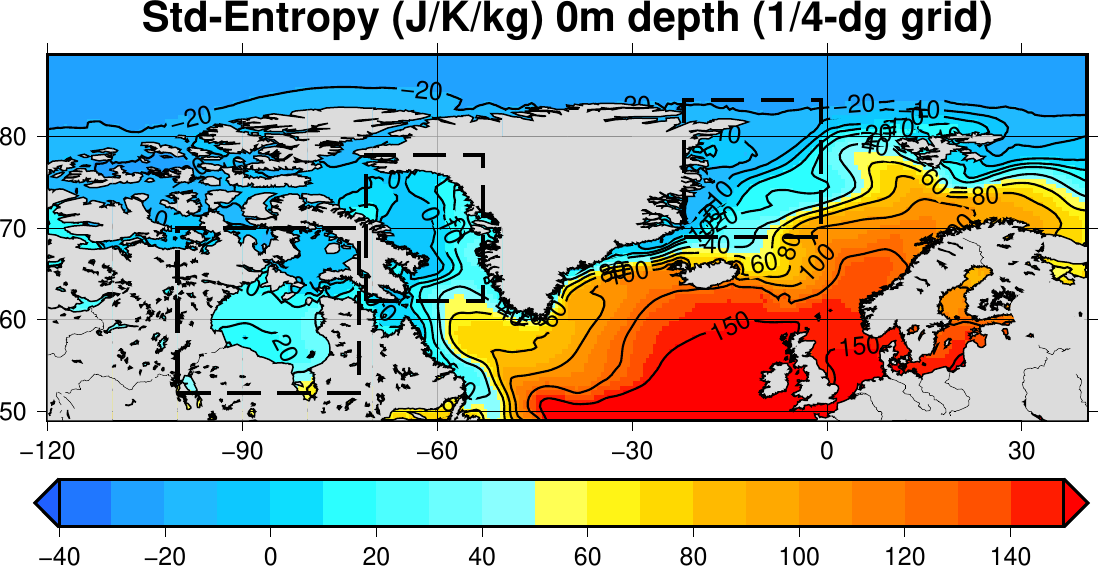}
\includegraphics[width=0.49\linewidth,angle=0,clip=true]{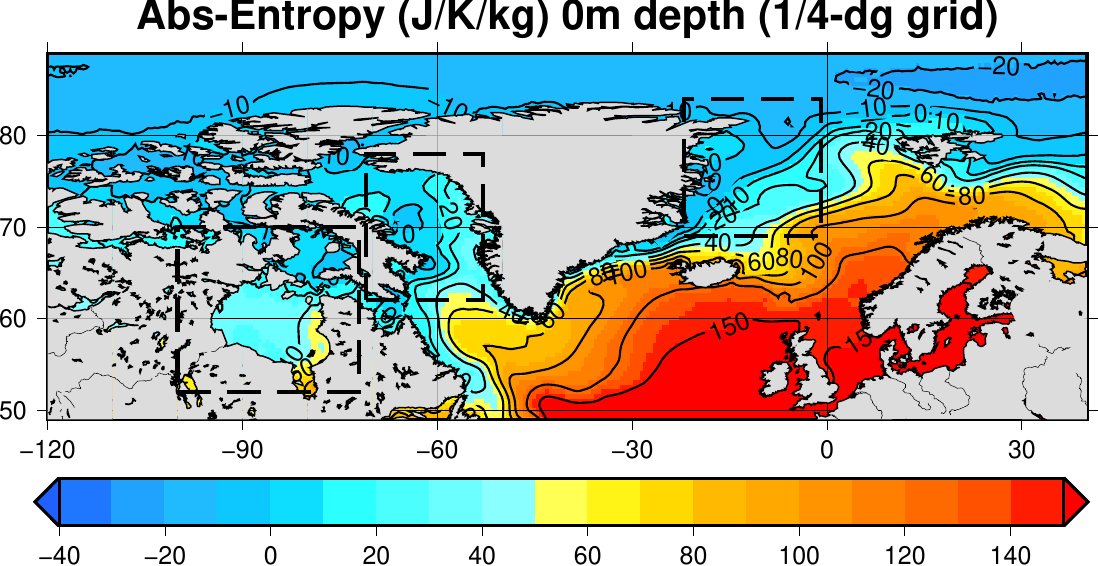}
\vspace*{2mm}\\
\caption{{\it \small The same as in the Figs.~\ref{Fig_WOA23_global}, but for the Arctic Ocean.  
\label{Fig_WOA23_regional_Arctic}}\\
-----------------------------------------------------------------------------------------------------------------------------------------------
}
\end{figure*}

\begin{figure*}[hbt]
\centering
\includegraphics[width=0.43\linewidth,angle=0,clip=true]{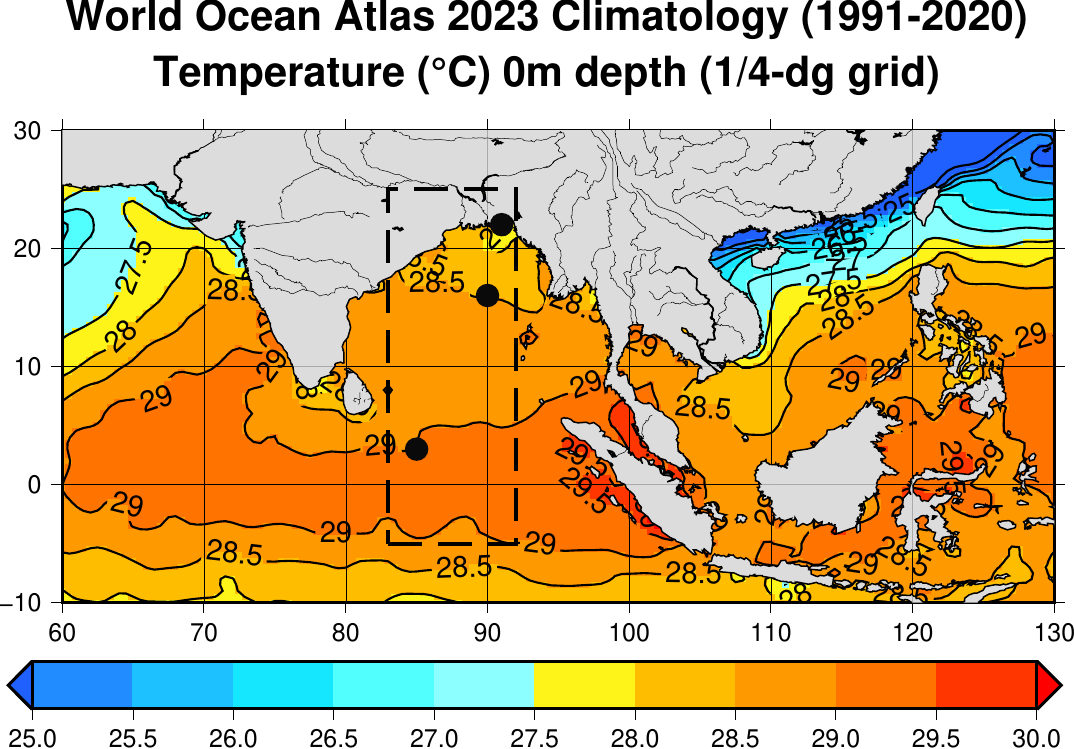}
\includegraphics[width=0.43\linewidth,angle=0,clip=true]{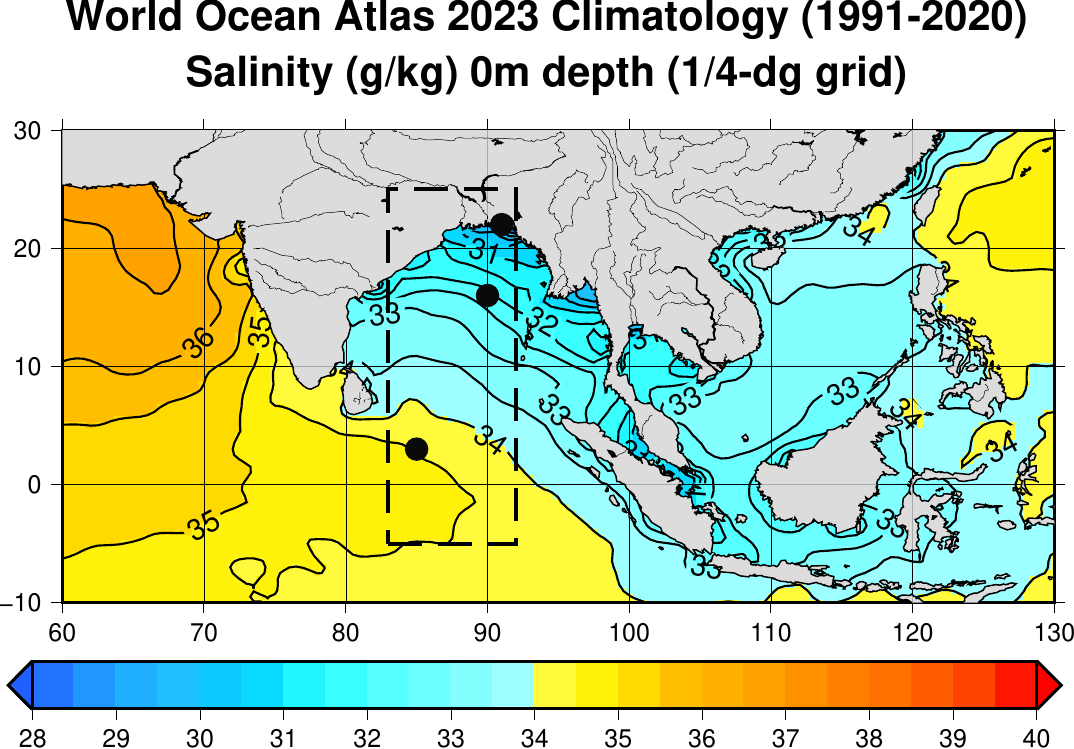}
\vspace*{2mm}\\
\includegraphics[width=0.43\linewidth,angle=0,clip=true]{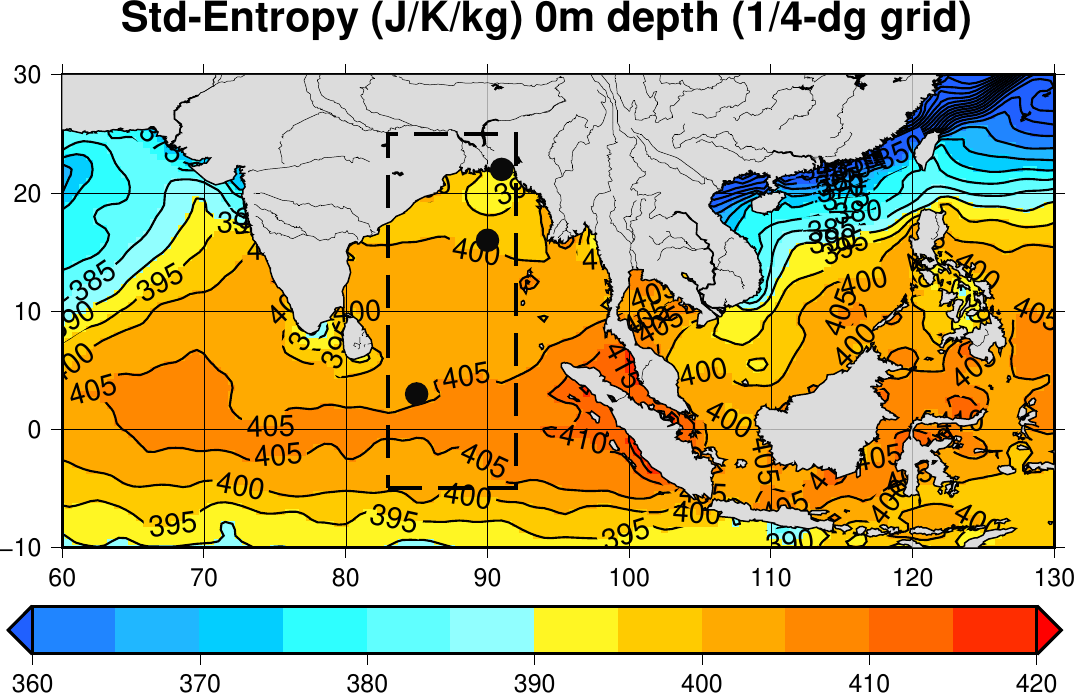}
\includegraphics[width=0.43\linewidth,angle=0,clip=true]{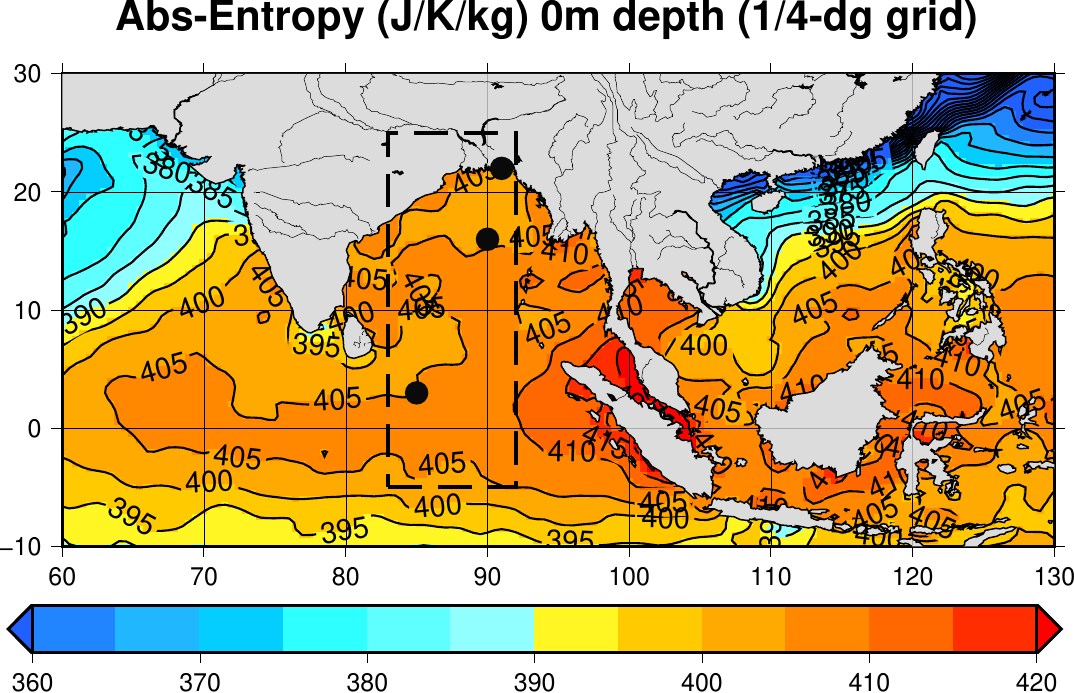}
\vspace*{2mm}\\
\includegraphics[width=0.4\linewidth,angle=0,clip=true]{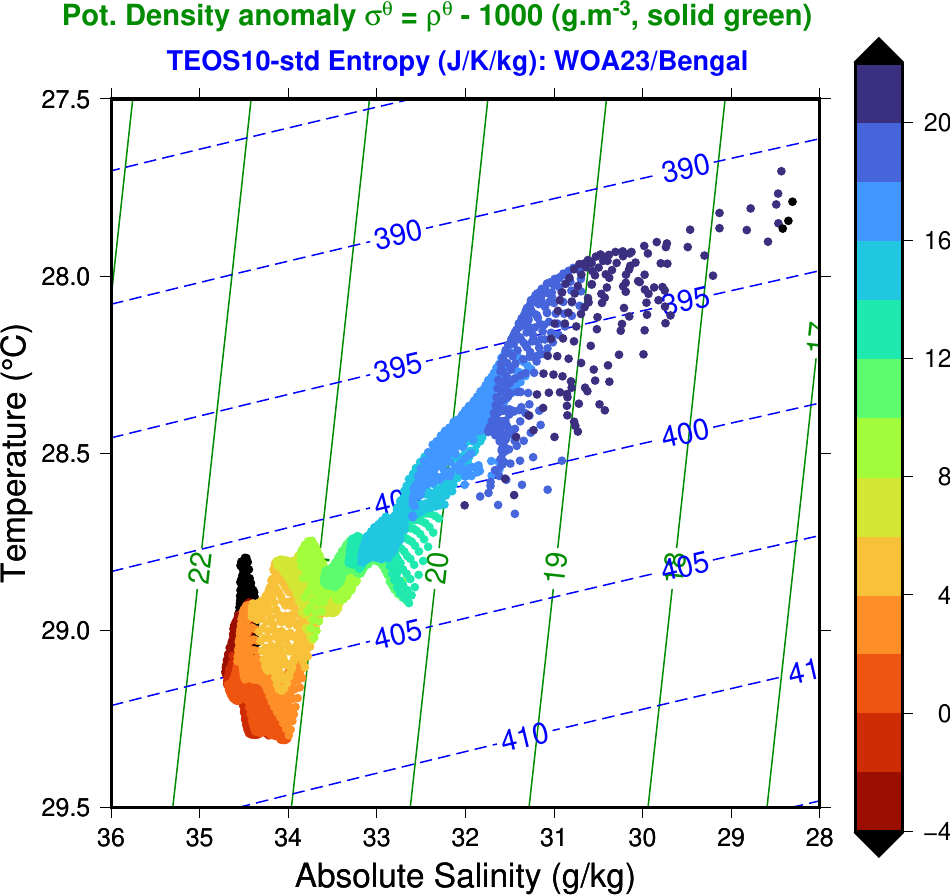}
\includegraphics[width=0.4\linewidth,angle=0,clip=true]{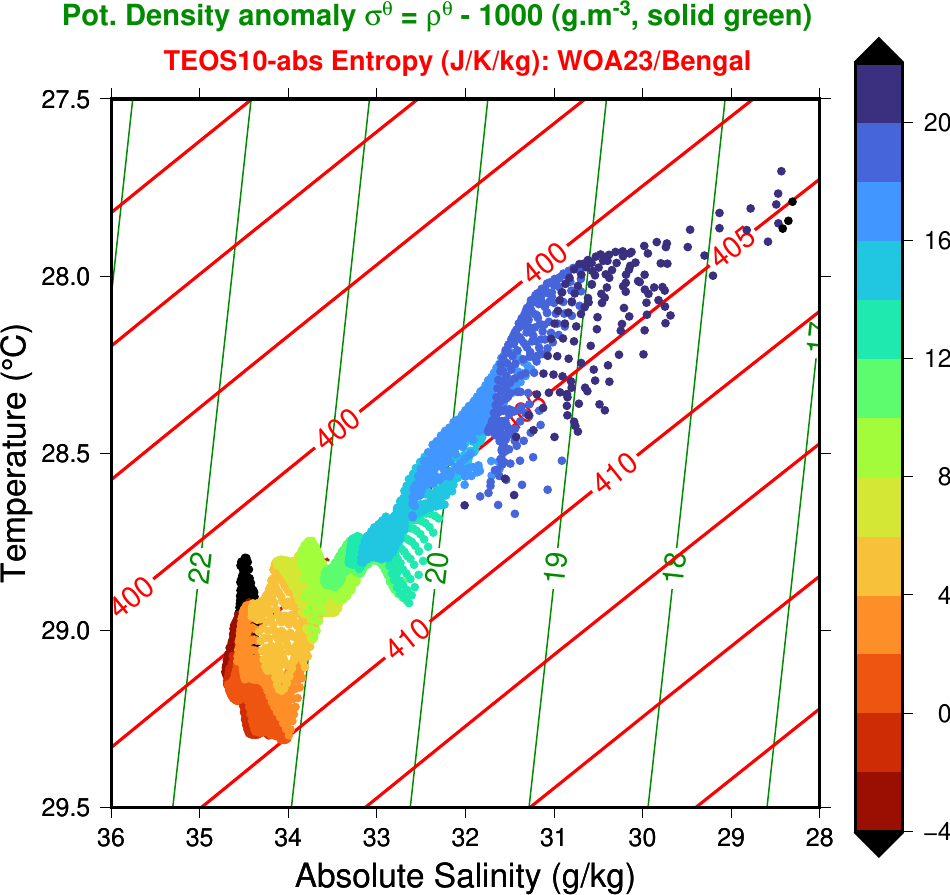}
\\
\caption{{\it\small The same as in Figs.~\ref{Fig_WOA23_global}, 
but for the Bay of Bengal and with three additional 
black disks at locations used to show numerical 
computations in the main text.
Two $t-S_{\rm A}$ diagrams are plotted with the TEOS 
standard-entropy blue-dashed lines (bottom left) 
and the absolute-entropy solid-red lines (bottom right).
The disks are coloured according to the latitude scale values
(from maroon for $4$~dg south to dark blue for $24$~dg north). 
Both salinity and temperature axes have been reversed, 
with decreasing values to the right and upward. 
\label{Fig_WOA23_regional_IndoAsie}}\\
-----------------------------------------------------------------------------------------------------------------------------------------------
}
\end{figure*}

The seawater standard (TEOS10) and absolute (third-law) entropies 
along the coasts of the Arctic Ocean are shown in the zoomed-in 
Figs.~\ref{Fig_WOA23_regional_Arctic}.

The blue ($<28$~g~kg${}^{\,-1}$) low values in salinity are due 
to the flow from the Mackenzie, Kolyma, Lena, Taz, and Ob rivers. 
The impacts of the associated large gradients in salinity correspond 
to more generalized (light-blue) absolute entropy values between 
$10$ and $30$ units off the coasts of Canada, Alaska, and Siberia, 
prolonging the same values off the west part of 
the Russian coasts (Barents and Kara Seas).

The surface values over the Arctic Ocean are thus more zonally symmetric 
with the absolute entropy than with the standard TEOS10 entropy. 
They are also more homogenized over the Bays of Hudson and Baffin, 
the Amundsen, Coronation, and Queen Maud Gulfs, and the Bering 
and Chukchi Seas. 
There are similarly more isentropic features in the Greenland Sea 
off the east coast of Greenland, where large gradients of 
salinity exist (see the dashed-black boxes).

The black line added to denote the Canada Basin (here at the longitude 
$140$~E/$220$~W) roughly corresponds to the SCICEX'97 Transect-A 
(longitude $150$~E/$210$~W), with the fresh water in the Beaufort Sea 
coming from the Mackenzie River.
The larger (light-blue) annual-mean values for the absolute entropy 
are in agreement with the surface values in the right part of 
the SCICEX'97 Transects (A and B) shown in 
Figs.~\ref{Fig_SCICEX97_transect_A} and \ref{Fig_SCICEX97_transect_B}.

\clearpage

\clearpage

\clearpage

 \subsection{A study of surface entropy: Bay of Bengal}
\label{subsection_global_surf_entropy_Bay_Bengal}
\vspace*{-2mm}

The seawater standard (TEOS10) and absolute (third-law) entropies 
over the Bay of Bengal are shown in the zoomed-in 
Figs.~\ref{Fig_WOA23_regional_IndoAsie}.

The low values of salinity ($<32$~g~kg${}^{\,-1}$, in blue) 
are due to the flow from the Ganges and Brahmaputra rivers 
and deltas, located at the north of this bay of Bengal, 
and also at the east of this bay from the Irrawaddy, 
Sittang, and Salween rivers.

The impacts of the low values and large salinity gradients over the Bay of Bengal, 
associated with low values but smaller temperature gradients, are to homogenise 
the seawater absolute entropy within the broad $405$-unit line 
(see the dashed boxes). 
The standard TEOS10 values exhibit larger north-south gradients with small 
values in the north of the Bay of Bengal and in the south with a narrow 
tongue formed by the $405$~unit line corresponding to the $29$°C 
temperature line. 

It is an interesting result to see how the waters to the north of 
the Bay of Bengal, with gradients in temperature and salinity, such as 
$27.7$°C and $27.8$~g~kg${}^{\,-1}$, 
$28.5$°C and $31.9$~g~kg${}^{\,-1}$, and  
$29.0$°C and $34.4$~g~kg${}^{\,-1}$,   
can produce such similar values of about 
$405 \pm 0.2$~J~K${}^{\,-1}$~kg${}^{\,-1}$ 
only for the absolute thermodynamic formulation of seawater entropy
(see the three black disks), and not for the present TEOS10's 
standard formulation of it, which varies by about $12$~units 
(namely $391.3$, $398.9$, and $403.5$~J~K${}^{\,-1}$~kg${}^{\,-1}$).
This cannot be due to chance, and it will be needed to better understand in the 
future which physical processes may be responsible for such a peculiar homogenisation 
of seawater absolute entropy, despite and because of such a special combination 
of temperature and salinity gradients. 
It should be pointed out that the rule (\ref{Eq_detas_dt_dSA}) for isentropic 
variations does not apply here, as this rule was only valid for temperatures 
close to zero Celsius, whereas here for $t \approx 28$°C all the terms 
of the $\eta(x,y,z)$ formulation that are non-linear in $y=t/(40) \approx 0.7$ 
should be taken into account.

It is also interesting to study the two $t-S_{\rm A}$ diagrams (see Part-I) 
shown at the bottom of Figs.~\ref{Fig_WOA23_regional_IndoAsie}, 
with the same green (isopycnic) iso-potential-density lines 
but with either the standard, arbitrarily modified TEOS10 version 
of the seawater entropy (on the left, with blue shaded lines) or 
with the absolute thermodynamic entropy (on the right, with solid red lines).
The seawater points plotted on these diagrams are those located within 
the domain ($-5$~dg/$+25$~dg) in latitude and ($83$~dg/$92$~dg) in longitude
(see the dashed boxes in the Figs.~\ref{Fig_WOA23_regional_IndoAsie}).

The absolute seawater entropy, in accordance with thermodynamics, shows 
(on the right) a remarkable but unexpected alignment of the points 
around the isentrope $405 \pm 3$~J~K${}^{\,-1}$~kg${}^{\,-1}$
for almost all the 
latitudes, despite significant gradients in temperature and salinity. 
In a different way, the blue curves (on the left) drawn with the standard 
TEOS10 formulation, which for this high range of temperature and salinity 
have a slope about three times lower than for the absolute entropy, 
do not provide any particular physical information and have a clear 
decorrelation with the scatterplot.

This confirms, for these tropical conditions, the importance of plotting 
these red absolute isentropic lines on the $t-S_{\rm A}$ diagrams, as 
already indicated when studying the polar SCICEX'96 and SCICEX'97 profiles.
This also means that, since the increased slope of these absolute isentropes 
compared to the TEOS10 formulation is a direct consequence of the addition 
of the absolute entropy increment 
$\Delta \eta_{\rm s} = (\eta_{\rm s0}-\eta_{\rm w0})(S_{\rm A}-S_{\rm SO})/1000$, 
which adds a linear function of the salinity depending on 
the reference values of the entropies $\eta_{\rm w0}$ of 
pure water and $\eta_{\rm s0}$ of sea salts, these values cannot 
be modified without physical consequences (as presently done 
with the TEOS10 formulation) on the calculation of the seawater 
entropy and second law.

 \subsection{A study of surface entropy: Northeast Atlantic and Mediterranean Seas}
\label{subsection_global_surf_entropy_Western_Europe}
\vspace*{-2mm}

\begin{figure*}[hbt]
\centering
\includegraphics[width=0.42\linewidth,angle=0,clip=true]{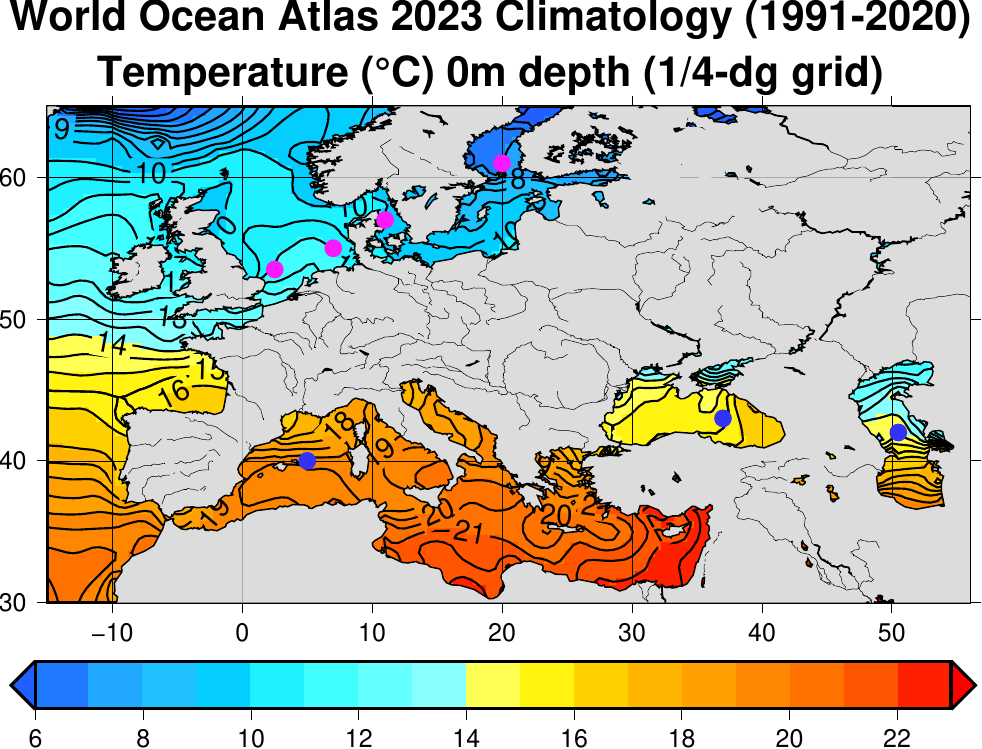}
\includegraphics[width=0.42\linewidth,angle=0,clip=true]{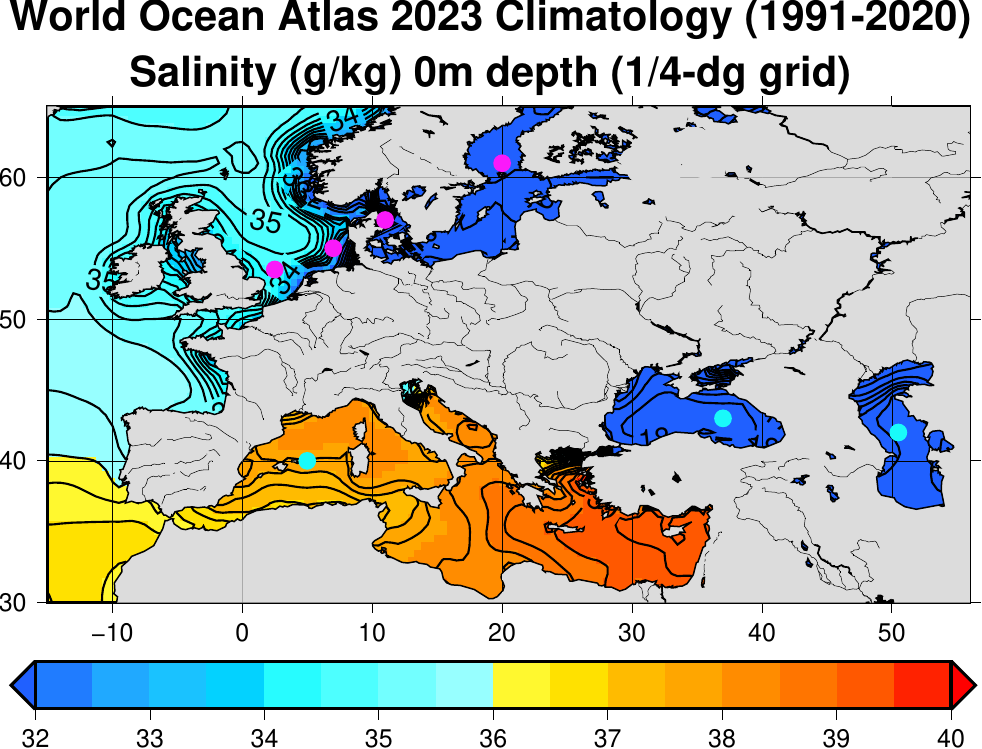}
\vspace*{2mm}\\
\includegraphics[width=0.42\linewidth,angle=0,clip=true]{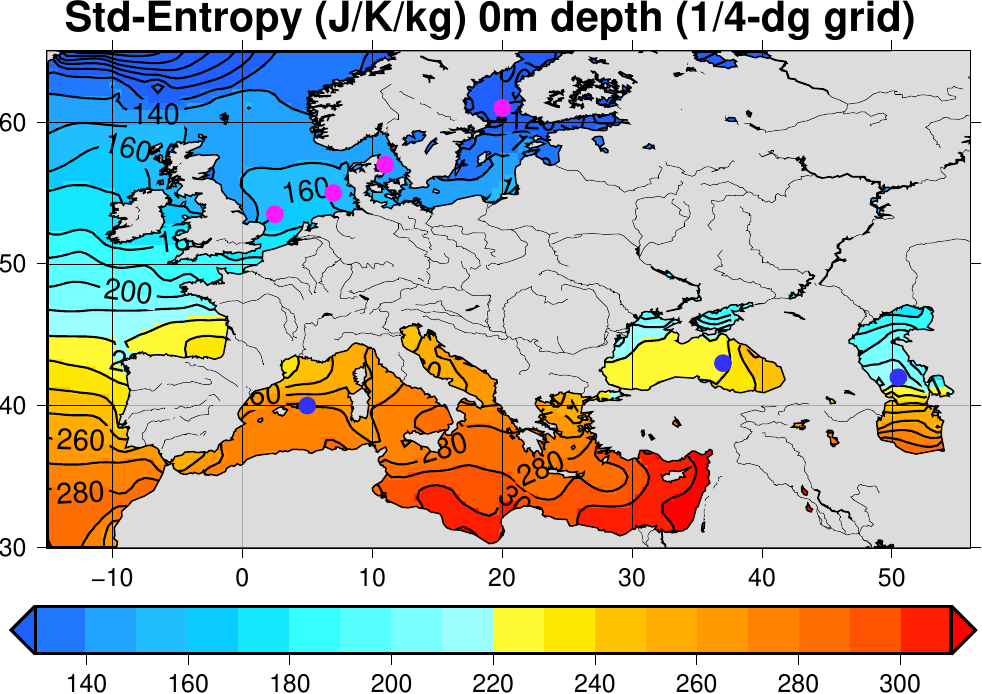}
\includegraphics[width=0.42\linewidth,angle=0,clip=true]{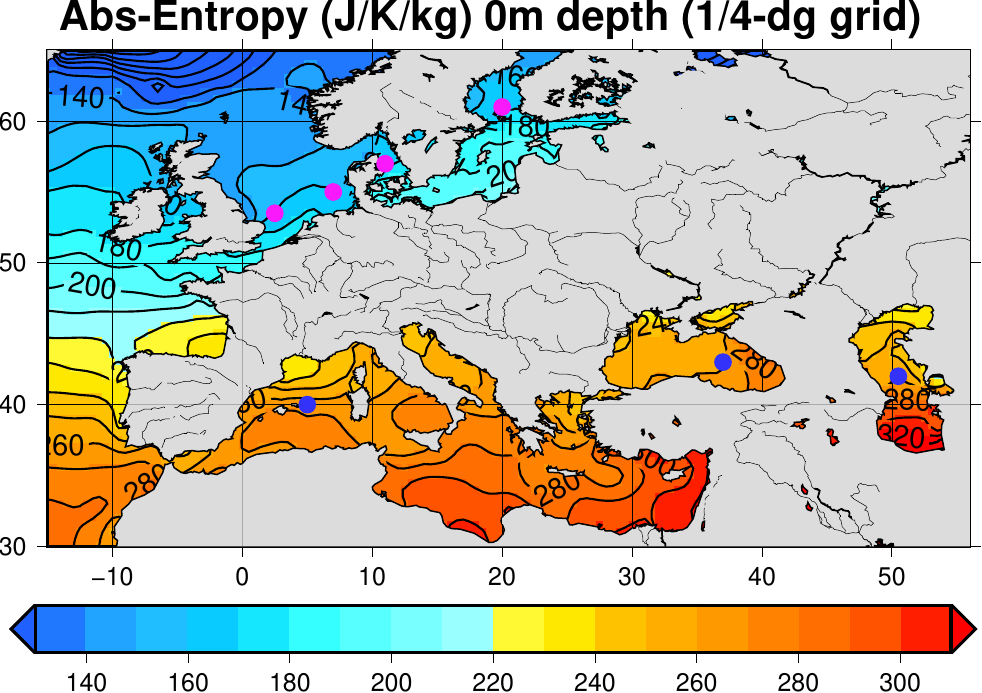}
\vspace*{2mm}\\
\includegraphics[width=0.4\linewidth,angle=0,clip=true]{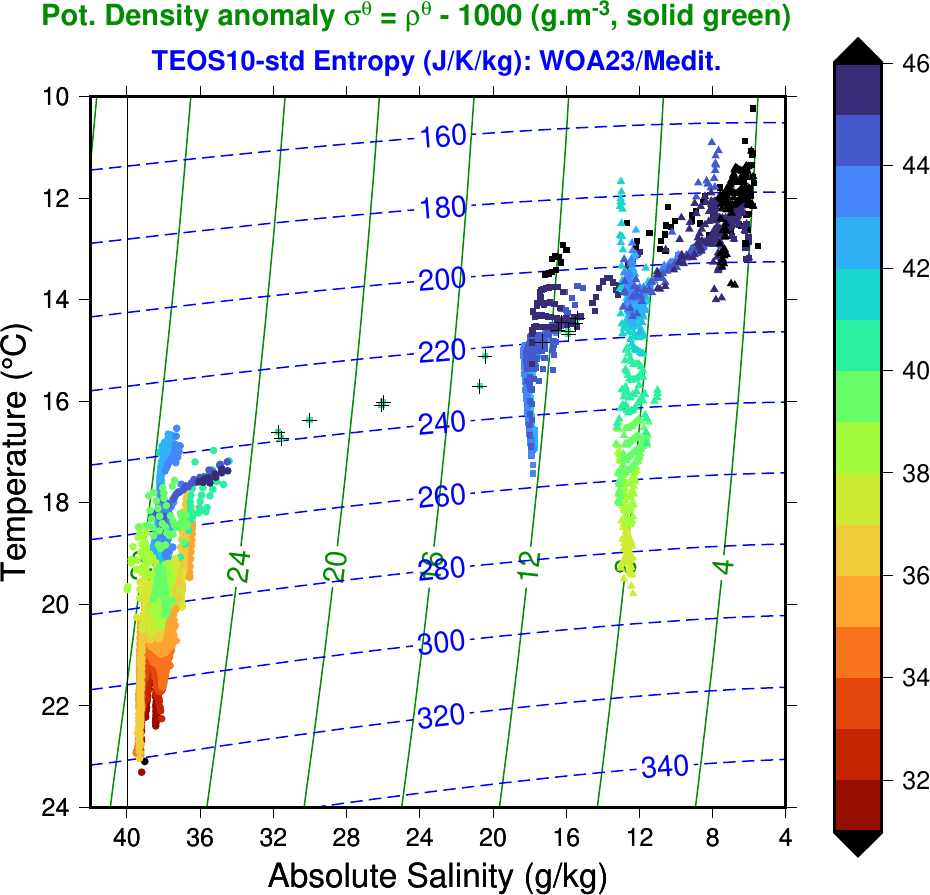}
\includegraphics[width=0.4\linewidth,angle=0,clip=true]{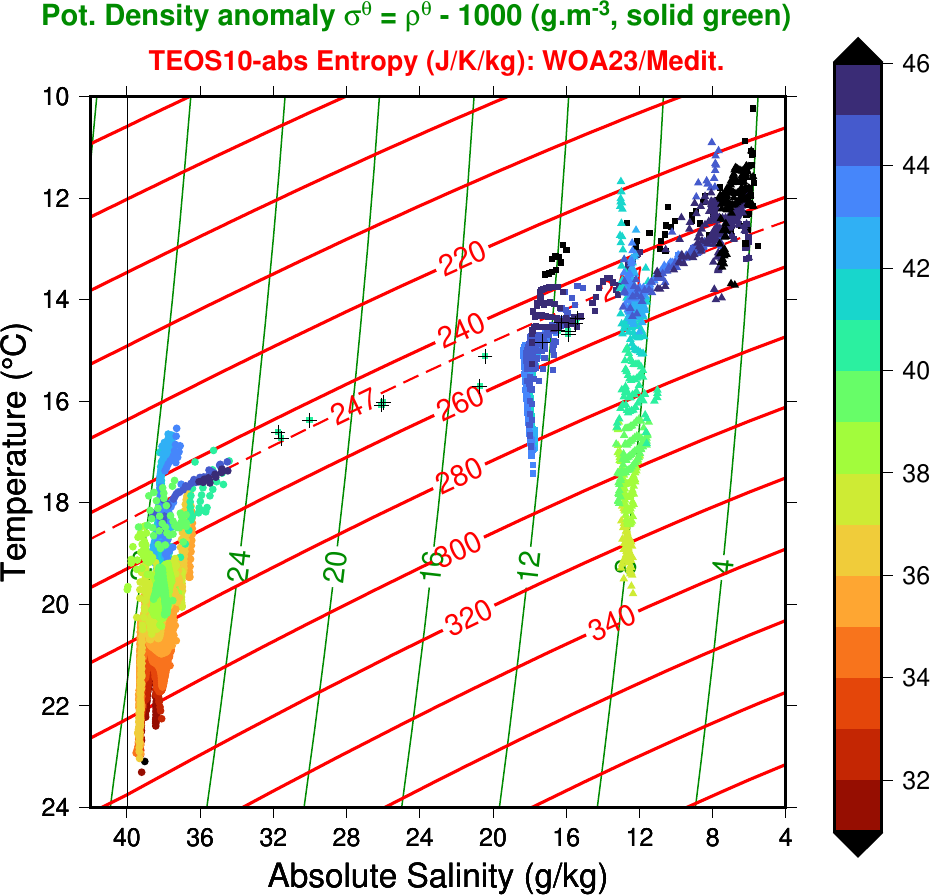}
\\
\caption{{\it\small The same as in the Figs.~\ref{Fig_WOA23_regional_IndoAsie}, 
but for the Western Europe and North Africa domain   
and with seven magenta and blue disks.
In the $t-S_{\rm A}$ diagrams,   
the Mediterranean Sea (discs) is to the left, 
the Caspian Sea (triangle) to the right, and 
the Sea of Marmara (crosses) and 
Black Sea (squares) in between.
\label{Fig_WOA23_regional_Mediter}}\\
-----------------------------------------------------------------------------------------------------------------------------------------------
}
\end{figure*}

The seawater standard (TEOS10) and absolute (third-law) entropies 
over western and central Europe, northern Africa, and the western 
part of the Middle East are shown in the zoomed-in 
Figs.~\ref{Fig_WOA23_regional_Mediter}.

In the Northeast Atlantic, colder temperatures with smaller salinity 
create more homogeneous absolute seawater entropy (light blue, 
between $170$ and $190$ units) over the Ireland Sea, off 
the whole coastal regions from south England and north 
of France (the English Channel), the south of the North Sea, 
Denmark (Skagerrak, Kattegat), the Baltic Sea, 
south of the Gulf of Bothnia, and up to the Gulf of Finland.

\clearpage

These absolute-isentropic features do not exist for the standard 
TEOS10 seawater entropy, which has almost exactly the same uneven 
shape as the temperature field and which is continuously 
decreasing from west to east off these coastal regions.
As an example (see the four magenta disks), 
the absolute entropy line $166 \pm 0.1$~J~K${}^{\,-1}$~J~kg${}^{\,-1}$ 
corresponds to very different $(t,S_{\rm A})$ conditions from west (North Sea) 
to east (Gulf of Bothnia):  
($11.5$°C, $34.5$~g~kg${}^{\,-1}$),  ($10.75$°C, $30$~g~kg${}^{\,-1}$), 
($10$°C,   $25.5$~g~kg${}^{\,-1}$), and ($7.2$°C, $5.9$~g~kg${}^{\,-1}$),
with a clear decrease in standard (TEOS10) entropy by more than $53$ units:
$164.8$, $156.5$, $147.7$, and $111.1$~J~K${}^{\,-1}$~J~kg${}^{\,-1}$, 
respectively.

Note that there is no point over the Russian Rybinsk reservoir 
and lakes of Ladoga and Onega for the WOA23 quarter-degree 
resolution dataset (accordingly, I have masked these 3 regions).

Over the Mediterranean Sea, the NW/SE gradient of temperature and 
W/E gradient of salinity combine with each other to create a 
slightly smoother and more N/S gradient in seawater absolute entropy.
The $260-270$~units (orange) band extends at almost constant latitude 
in the whole Mediterranean Sea (from Barcelona, Naples, Athens, and 
Istanbul). 
There is, moreover, an unexpected extent toward the Black and Caspian 
Seas, which have especially low values of salinity and similar 
absolute seawater entropy as in the Mediterranean Sea. 
These features do not exist for the standard (TEOS10) seawater 
entropy, which again has almost exactly the same uneven shape as 
the temperature field, and thus with less homogeneous values for the 
Mediterranean, Black, and Caspian Seas. 

Note that there is no point over the Aral Sea for the WOA23 
quarter-degree resolution dataset (not shown).
Note also that the relevance of the TEOS10 formulation to 
compute the entropy of low-salinity lakes (like here for 
the Caspian Sea) has been recalled by 
\citet[][p.477]{Feistel_2018}, who stated that: 
``{\it\,based on the heuristic, empirical ``Millero rule'', which argues 
``\,that the physical chemical properties of most lakes can be 
determined from equations for seawater at the same total salinity\,''
(...) the properties of anomalous seawater can be estimated
by the TEOS-10 functions in terms of SSW properties evaluated 
at the same density, temperature, and pressure.\,}''

To confirm and better illustrate these unexpected and astonishing 
almost absolute-isentropic states for these three Mediterranean, 
Black, and Caspian Seas, the three typical conditions 
($19.0$°C, $38.3$~g~kg${}^{\,-1}$), 
($15.4$°C, $18.0$~g~kg${}^{\,-1}$), and 
($14.5$°C, $12.2$~g~kg${}^{\,-1}$), 
corresponding to the three (east/west) blue disks in 
Figs.~\ref{Fig_WOA23_regional_Mediter}, 
lead to similar values of absolute entropy 
($260 \pm 0.2$~J~K${}^{\,-1}$~kg${}^{\,-1}$).  
Differently, the standard (TEOS10) entropies are decreasing
by about $50$~units 
($266.3$, $227.5$, and $216.7$~J~K${}^{\,-1}$~kg${}^{\,-1}$). 

As with the Bay of Bengal (Figs.~\ref{Fig_WOA23_regional_IndoAsie}), 
it is also useful to study the $t-S_{\rm A}$ diagrams shown at the bottom 
of Figs.~\ref{Fig_WOA23_regional_Mediter} with points 
coloured by latitudes (from maroon for $31$~dg north to 
dark blue for $46$~dg north) for  
the Mediterranean Sea (discs), 
the Black Sea (squares), 
the Caspian Sea (triangles), and 
the Sea of Marmara (crosses).
The smaller latitude points are roughly aligned
along isopycnic green lines:  
the Mediterranean Sea has the higher density 
($26$ to $30$~g~kg${}^{\,-1}$), 
the Caspian Sea the smaller density 
($8$ to $9$~g~kg${}^{\,-1}$), with the 
Black Sea in between
($12$ to $13$~g~kg${}^{\,-1}$)
and the Sea of Marmara connecting the 
Mediterranean and Black Seas (rapid variations 
between $24$ and $12$~g~kg${}^{\,-1}$).

Clearly the standard values of the TEOS10 entropy 
(blue dashed lines, on the left) are not correlated 
with the Mediterranean, Marmara, Caspian, and Black
Sea scatterplots.
Differently, the new absolute seawater entropy red
lines (on the right) roughly connect the points 
at the same latitude and colours.
Moreover, there are unexpected absolute isentropic rough 
alignments of the higher (northern, blue, and dark-blue) 
symbols along the $247$ absolute entropy dashed-red line,
including the connecting Marmara's crosses.
Since the red absolute entropy lines are unknown to the 
WOA23 objective analysis process, this means that nature 
has chosen to follow either the isopycnic green lines 
(for the lowest latitudes) or the red absolute seawater 
entropy lines (for the highest latitudes), with 
no bias due to the WOA23 processes acting
separately on $t$ and $S_{\rm A}$. 

As with the Bengal Sea and the absolute entropy of the atmosphere, 
this means, on the one hand, that the absolute entropy defined 
in thermodynamics has a hidden physical meaning, which is revealed 
here by the $t-S_{\rm A}$ oceanic diagrams. 
On the other hand, it is not possible to arbitrarily modify the 
increment $\Delta \eta_{\rm s}$ by freeing oneself from the use of 
the reference values of entropies, as recommended by the third law 
and provided in all the Tables of Thermodynamics.
  
Note that the more zonally symmetric feature of the 
$240-260$~entropy-unit band cannot be explained by pure 
oceanic processes, simply because the Caspian Sea does not 
communicate with the two other seas, the Black and Mediterranean Seas.
Therefore, it will be needed to better understand in the future 
which physical processes may be responsible for such a peculiar 
homogenisation of absolute seawater entropy. For instance, this 
may be due to more general and zonally symmetric atmospheric processes
and via the interface between the atmosphere and the ocean.

 \section{Conclusion}
\label{section_conclusion}
\vspace*{-2mm}

In this Part~II, the absolute seawater entropy $\eta_{\rm abs}$ defined 
in Part~I is studied for the first time with concrete case studies and 
applications to oceanic vertical CTD profiles and analysed surface datasets.
The absolute entropy $\eta_{\rm abs}$ recalled in (\ref{Eq_Delta_eta_ans_std}) 
is computed by adding the salinity increment $\Delta \eta_{\rm s}$ recalled 
in (\ref{Eq_etas_minus_etaw_value}) to the standard TEOS10 value 
$\eta_{\rm std/TEOS10}$ recalled in Part~I.

The case studies undertaken in this Part~II reveal new homogeneous 
and isentropic features, both vertically and horizontally at the surface, 
with the property that only the absolute version of the seawater entropy 
can reveal these new isentropic regions. 

These peculiar absolute-isentropic features are clearly visible on 
the three oceanic $t-S_{\rm A}$ diagrams studied for polar (SCICEX'96 and 97), 
tropical (Bay of Bengal), and mid-latitude (Mediterranean, Marmara, 
Black, and Caspian Seas) conditions.
The scatterplots appear 
to preferentially follow either iso-potential-density green lines  
or iso-absolute-entropy red lines, to the exclusion of the blue lines 
and the ``arbitrarily modified seawater entropy'' presently offered
as an output of TEOS10.
These results could not have been obtained by randomly juxtaposing 
oceanic states with very different temperatures and salinities 
and leading (by chance) to the same absolute-entropy values. 

Generally speaking, it was worth studying the impact of the increment term 
$\Delta \eta_{\rm s}$ wherever salinity values are particularly low or high, 
but also wherever salinity gradients are large.
It can even be said, insofar as the TEOS10 formulation leads to almost 
the same spatial structures as for temperature, that there is very little 
advantage in proposing the current standard entropy field as an output 
of TEOS10.

Differently, the absolute entropy field, including the thermodynamically 
consistent increment term $\Delta \eta_{\rm s}$, seems to provide a real 
added value, which should justify its calculation and its availability 
as an output of TEOS10.
But there is nothing surprising in this, since absolute entropy is one 
of the fundamental quantities of thermodynamics (along with temperature 
and energy), and only the inclusion of this increment $\Delta \eta_{\rm s}$ 
makes it possible to calculate this entropy rigorously for the seawater, 
as prescribed by the third law of thermodynamics, which cannot be called 
into question with impunity (as done in TEOS10).

Possible physical explanations for these alignments along absolute 
isentropes may include turbulent processes occurring in the atmosphere, 
in the oceans, and at the interfaces between the two.
Indeed, partly suggested by my earlier studies of the absolute entropy 
of the atmosphere, this uniformity of absolute seawater entropy, 
and therefore the isentropic status of surface waters, 
may be due to atmospheric turbulent processes following the recommendations 
of \citet[][]{Richardson_19a,Richardson_22}., 
He explained that 
the atmospheric turbulence must involve the wind components, 
the total water content, and the absolute entropy 
(or the associated potential temperature).
In line with this hypothesis, I have shown that turbulent schemes 
are preferentially applicable to the absolute entropy 
for the atmosphere 
\citep{Marquet2011QJ,MarquetBechtold2020}  
and for oceanic bulk schemes 
\citep{MarquetBelamari2017WGNELewis}, 
with in all probability the same turbulent processes that 
should be applied to the absolute seawater entropy 
(and not to the temperature $t$ nor the actual potential 
temperatures $\theta$ or $\Theta$).

Therefore, a possible interpretation for these isentropic oceanic structures 
may be found in the more homogeneous and zonally symmetric atmospheric 
circulation and in the corresponding impacts on the oceanic state variables 
via the bulk-surface turbulent processes acting on the absolute entropy.
Note, however, that the turbulence processes applied to absolute entropy  
do not prevent but add to the impacts linked to convective phenomena, 
which are linked to buoyancy and density-related variables (such as virtual 
temperature for the atmosphere and isopycnic lines for the ocean).
It could therefore be imagined a sort of competition between these two 
turbulent and convective effects, acting on different variables, 
as shown by the scatterplots on the $t-S_{\rm A}$ diagrams, which follow 
one or the other of the families of green (isopycnic) or red 
(absolute-isentropic) curves.

It should be emphasized that the correction term $\Delta \eta_{\rm s}$ depends 
linearly on salinity and thus varies not only spatially but temporally, which 
fundamentally changes the  calculation of entropy itself beyond the addition 
of a simple constant (like $\eta_{\rm w0}$ in $\eta^{\rm W}_{\rm Fei03}$ recalled 
in Table~1 of Part~I). 
Consequently, studies of the type ``\,maximum entropy state\,'' defined by 
$d\eta = 0$ depend on a certain combination of the differentials $dT$ and 
$(\eta_{\rm w0}-\eta_{\rm s0})\:dS_{\rm A}$. 
There is, therefore, an impact of the absolute 
reference entropy terms $\eta_{\rm w0}$ and $\eta_{\rm s0}$ for non-stationary 
states with both $dT \neq 0$ and $dS_{\rm A} \neq 0$, 
as, for example, for the regional, seasonal, and climate changes. 

Furthermore, insofar as the entropy gradient regions influence the turbulent transport of this 
thermodynamic state variable, with turbulent flows that must cancel out in isentropic 
regions, the absolute definition of entropy must be preferred to any other arbitrarily 
modified definition; otherwise, the entropy budget and therefore the second principle 
of thermodynamics will be arbitrarily modified.
In fact, the entropy balance equation cannot be reduced to the positive production 
term usually studied, and entropy flows (often discarded) at the ocean-atmosphere 
interface depend on $\Delta \eta_{\rm s}$ and on the absolute definition of 
entropy studied in the paper.

This study should be continued in order to better understand the origin of the zonally 
symmetric and local three-dimensional oceanic structures of the absolute seawater entropy,  
in particular in the Bay of Bengal, in the Mediterranean, Black, and Caspian Seas,
the Northeast Atlantic and Mediterranean Seas, 
as well as in the whole Arctic Ocean.


\section*{Acknowledgments}
\vspace*{-3mm}

The author would like to thank the anonymous referees and the editor
for the constructive comments, which help to improve the manuscript. 

I would also like to extend my warmest thanks to Jean-Claude Andr\'e, 
whose invaluable and crucial assistance made it possible to publish 
both parts of this paper.

The Fig.~1 by \citet[][p.2]{Steele_al_JGR_2004} is 
free of Copyright by the AGU  (J. Geophys. Res. Oceans).
The SCICEX'96 and 97 CTD vertical profiles 
and SCICEX'97 Sample Locations map 
have been downloaded in September 2024 at:
\url{https://nsidc.org/scicex} and  
\url{https://www.nodc.noaa.gov/archive/arc0021/0000568/1.1/data/0-data/}, 
see the Supplementary Data. 

The World Ocean Atlas 2023 climatology 
(WOA23, release February 2024, quarter-degree grid) 
are freely available at:
\url{https://www.ncei.noaa.gov/products/world-ocean-atlas}.

The TEOS10-GSW Oceanographic FORTRAN software has been downloaded from 
\url{https://www.teos-10.org/software.htm} 
\citep{McDougall_TEOS10_GSW_Getting_Started_2011}.


The SCICEX and WOA23 directories were broken in October 2024 due to Hurricane Helene Outage.
I have made available in the Zenodo site \citep{Marquet_SCICEX_2024a,Marquet_SCICEX_2024b} 
the SCICEX (96 and 97) CTD and WOA23 public sub-datasets used to plot the figures~1 to 7.

First-Review Answers to the Editors and Reviewers 
can be found on Zenodo \citep{Marquet_Zenodo_2025_Answers_R1}.

Supplementary materials are provided in the Zenodo file
\citet{Marquet_Zenodo_2025_Sup_Mat_3rd_law}. 
 
Second-Review Answers to the Reviewers 
can be found on Zenodo \citep{Marquet_Zenodo_2026_Answers_R2}.

\bibliographystyle{ametsoc2014}
\bibliography{Marquet_seawater_absolute_entropy_Part2_arXiv_R2}

\clearpage

\newpage


 \end{document}